\documentclass[%
  twocolumn,
superscriptaddress,
 amsmath,amssymb,
 aps, prb,
]{revtex4-2}

\usepackage{tikz}
\usepackage{braket}
\usepackage{siunitx}
\usepackage{xcolor}
\usepackage[toc]{glossaries}
\usepackage{graphicx}
\usepackage{dcolumn}
\usepackage{bm}
\usepackage[colorlinks=true,
                linkcolor=blue,
	        citecolor=blue,
	        urlcolor=blue]{hyperref}
\usepackage[normalem]{ulem}

\usepackage[colorlinks=true,
                linkcolor=blue,
	        citecolor=blue,
	        urlcolor=blue]{hyperref}

\newcommand{\trieste}{Dipartimento di Fisica, Universit\`a di Trieste,  I-34151 Trieste, Italy}
\newcommand{\sissa}{Scuola Internazionale Superiore di Studi Avanzati (SISSA), I-34136 Trieste, Italy}

\begin{document}

\title{AiiDA-TrainsPot: Towards automated training of neural-network interatomic potentials}%

\author{Davide Bidoggia\footnotemark[1]}
\thanks{These authors contributed equally to this work.}
\affiliation{\trieste}

\author{Nataliia Manko\footnotemark[1]}
\thanks{These authors contributed equally to this work.}
\affiliation{\trieste}
\affiliation{\sissa}

\author{Maria Peressi}
\affiliation{\trieste}

\author{Antimo Marrazzo}
\email{amarrazz@sissa.it}
\affiliation{\sissa}

\date{\today}

\begin{abstract}
Crafting neural-network interatomic potentials (NNIPs) remains a complex task, demanding specialized expertise in both machine learning and electronic-structure calculations. Here, we introduce AiiDA-TrainsPot, an automated, open-source, and user-friendly workflow that streamlines the creation of accurate NNIPs by orchestrating density-functional-theory calculations, data augmentation strategies, and classical molecular dynamics.
Our active-learning strategy leverages on-the-fly calibration of committee disagreement against \emph{ab initio} reference errors to ensure reliable uncertainty estimates. We use electronic-structure descriptors and dimensionality reduction to analyze the efficiency of this calibrated criterion, and show that it minimizes both false positives and false negatives when deciding what to compute from first principles. AiiDA-TrainsPot has a modular design that supports multiple NNIP backends, enabling both the training of NNIPs from scratch and the fine-tuning of foundation models. We demonstrate its capabilities through automated training campaigns targeting pristine and defective carbon allotropes, including amorphous carbon, as well as structural phase transitions in monolayer $\mathrm{W_xMo_{1-x}Te_2}$ alloys. 
\end{abstract}

\maketitle

\section{Introduction}
In computational materials science, many key properties\textemdash such as phase transitions, diffusion, viscosity, thermal transport\textemdash arise from atomic and molecular motion over time. Chemical reactions are typically rare events, requiring long simulations times and enhanced sampling techniques. Because these properties emerge from dynamic processes rather than static configurations, molecular dynamics (MD) simulations are essential for capturing their time-dependent behavior and understanding material performance under different conditions. MD simulations enable the study of crucial phenomena in materials and molecular systems, such as heat conduction, ion transport in batteries, crack propagation in materials, just to name a few, and the behavior of soft matter and biomaterials under different conditions. Fully \emph{ab initio} molecular dynamics (AIMD) simulations can be highly accurate but computationally expensive, as they require solving the Schrödinger equation every timestep to determine atomic forces. This makes AIMD impractical for large systems or long timescales. On the other hand, empirical interatomic potentials, which approximate atomic interactions with predefined functional forms, are computationally efficient but often lack the accuracy needed for quantum mechanical effects. Additionally, these potentials are typically parameterized for specific materials and struggle to generalize across diverse chemical environments, limiting their predictive power for complex or novel systems. For a long time, the Car-Parrinello method (CPMD)~\cite{CP_prl_1985} offered a compromise between AIMD and empirical potentials by evolving both electronic and ionic degrees of freedom simultaneously with a Lagrangian formalism, reducing the need for explicit electronic structure calculations at every step. While CPMD improves efficiency compared to traditional AIMD, it still remains computationally demanding and is limited by the need for careful tuning of fictitious electron mass parameters, which can affect accuracy and stability.

Machine learning (ML) and neural-network interatomic potentials (NNIPs) have revolutionized the field, by bridging the gap between \emph{ab initio} accuracy and computational efficiency~\cite{ Deringer2019}. These models are trained on high-fidelity density functional theory (DFT) data, allowing them to capture complex quantum mechanical interactions with far greater accuracy than traditional empirical potentials. Unlike fixed functional forms, ML potentials can flexibly generalize to diverse chemical environments while maintaining computational costs significantly lower than AIMD. This breakthrough has enabled large-scale and long-timescale simulations of materials with near-quantum accuracy, essentially extending the scope of what can be simulated from first principles.

Over the last years, accuracy and data efficiency of ML interatomic potentials have improved remarkably, often at the price of increased algorithmic complexity. In this context, NNIPs based on equivariance have emerged as particularly promising and several architectures have been proposed, including NequIP~\cite{Batzner2022}, Allegro~\cite{Musaelian2023}, MACE~\cite{MACE} and Point Edge Transformer (PET)~\cite{pet}. More recently, foundation models leveraging large-scale pretraining on diverse chemical datasets, provided a path towards improved transferability, data efficiency, and accuracy across a wide range of materials and molecular systems~\cite{maceoff_arxiv, petmad}.

As of today, the training of an accurate NNIP remains a complex and time-consuming task. First, high-quality NNIPs require training datasets of the order of thousands of supercell single-point \emph{ab initio} calculations with hundreds of atoms in the unit cell; millions if foundation models for the entire periodic table are targeted. Most notably, the accuracy and extrapolation capabilities of NNIPs hinge on the careful choice of the training structures, which have to be sufficiently diverse and abundant to avoid overfitting. While foundation models promise transferability across chemical space, fine-tuning them to high accuracy for a given material family still requires curated datasets.

In this work, we introduce AiiDA-TrainsPot, an automated, open-source and user-friendly framework designed to automate the training of NNIPs. Our approach integrates automated workflows for DFT calculations with neural-network training and classical MD to systematically explore the potential energy landscape through random distortions, strain, interfaces, neutral vacancies, trajectories at varying temperatures and pressures.  Existing platforms such as DP-GEN~\cite{Zhang2020} or SchNetPack~\cite{Schutt_JCTC_2018} have demonstrated the power of active learning for NNIPs. However, most existing frameworks are tied to specific NNIP architectures or training backends, offer only limited error handling for first-principles simulations, and do not provide systematic provenance tracking of the training process. In contrast, AiiDA-TrainsPot offers (i) full code-agnostic modularity across quantum engines, ML architectures, and MD codes; (ii) an extensive suite of automated dataset augmentation strategies (defects, slabs, clusters, substitutions, alloys); and (iii) a \emph{calibrated} committee-disagreement scheme that provides quantitative uncertainty estimates even in production runs.  The method is validated through the fully automated training and fine-tuning of NNIPs for a diverse range of carbon allotropes, including amorphous carbon, as well as for monolayer $\mathrm{W_xMo_{1-x}Te_2}$ alloys, achieving state-of-the-art accuracy and data efficiency. Combined with AiiDA’s reproducibility infrastructure, these features uniquely position AiiDA-TrainsPot as both a democratizing tool for domain scientists and a robust platform for future foundation-model development. 

\section{Results and discussion}

\begin{figure*}[tbp]
  \centering

   \includegraphics[trim={0 0 0 0px},clip,width=0.9\textwidth]{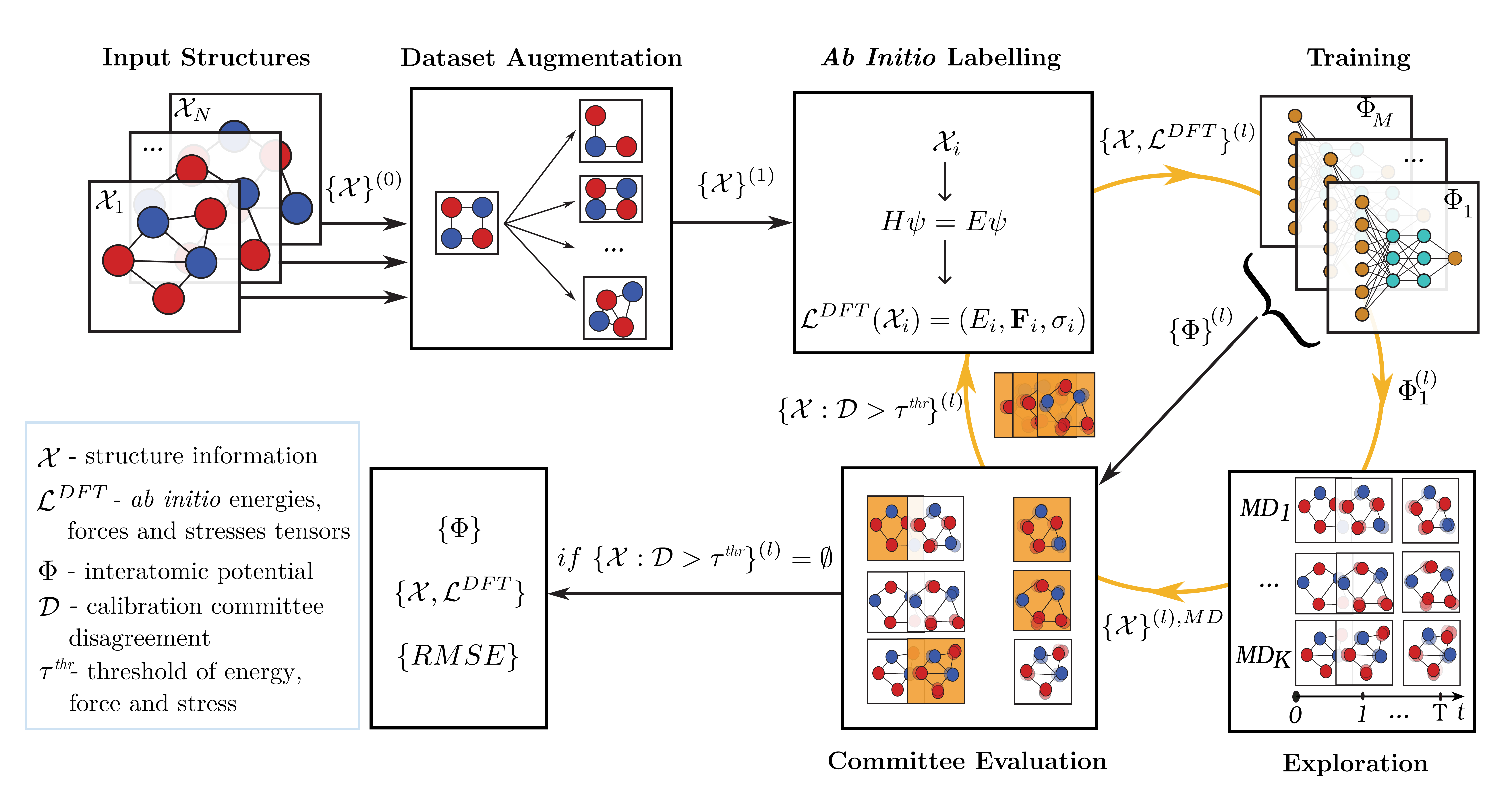}
\caption{\label{Fig1}{\bf Schematic representation of the AiiDA-TrainsPot automated workflow.} Initial input structures (\{$\mathcal{X}\}^{(0)}$) can be augmented by creating configurations with random distortions, strain, vacancies, cluster and slab extraction.
\textit{Ab initio} calculations are performed on these configurations (\{$\mathcal{X}_i$\}) that are thus labeled by energies, forces, and stress tensors which constitute the \textit{reference} data  (\{$\mathcal{L}_i$\}). A committee of potentials (\{$\Phi_j$\}) is trained on these configurations and committee evaluation is used to compare their predictions ($\mathcal{L}_k(\Phi_j)$) on structures  ($\mathcal{X}_k$) that are extracted from MD trajectories at different temperatures and pressures obtained through one of the potentials of the committee ($\Phi_1$). If for a structure $\mathcal{X}_k$  the disagreement ($\mathcal{D}_k$) averaged over the committee exceeds a threshold, the structure is added to the \textit{reference} training dataset and the  \textit{ab initio} labelling is performed. This iterative process continues until convergence is achieved and workflow outputs labeled structures, trained potentials, and root mean square errors (RMSE) for energies, forces, and stress tensor components. }

\end{figure*}

\subsection{Automation strategy}
AiiDA-TrainsPot is based on a two-stage augmentation process shown in Fig.~\ref{Fig1}. In a typical use case, users provide a handful of structures\textemdash from one up to tens\textemdash that are augmented to the order of thousands through structural manipulations. This happens in the first stage, where about a thousand of structures are calculated with \emph{ab initio} methods (here DFT) and used to train the first generation of NNIPs. The second stage employs MD simulations at different thermodynamic conditions to accurately sample the basins of the potential energy surface (PES), thus exploring regions that are particularly relevant for applications. All MD trajectories are obtained with the NNIPs trained in the previous workflow step. Some structural configurations are sampled from MD trajectories and calculated \emph{ab initio} to train a second generation of NNIPs. The choice of the structures to label with \emph{ab initio} results is based on the committee disagreement, i.e., the spread in the predictions of a committee of NNIPs initialized with different seeds and trained on the same data. At each iteration a new generation of potentials is trained and the committee disagreement is calibrated on actual deviations between NNIP predictions and the baseline level of theory, i.e., DFT.

The active learning loop continues until errors on energy, forces and stress tensor are below a user-defined threshold. We emphasize that AiiDA-TrainsPot supports multiple use cases depending on the available input data, which goes beyond the generation of a NNIP from scratch and includes the fine-tuning of foundation models (see Sec. \ref{sec:code}). In the following, we discuss in detail each step of the workflow.

\subsubsection{Input structures}
AiiDA-TrainsPot can start from a small set of initial atomistic structures $\left\{\mathcal{X}\right\}^{(0)}$, determined by boundary conditions (periodic vs. open), cell parameters, atomic species and atomic positions. The number and diversity of input structures should reflect the target applications: for example, the study of temperature-dependent properties of diamond might require a single input structure, the development of NNIP for all carbon allotropes would probably include at least all known crystalline prototypes of carbon, while universal (a.k.a. foundation) models for the entire periodic table might require tens of thousands of input structures, which could be obtained from computational materials databases such as  Materials Cloud~\cite{MaterialsCloud}, Materials Project~\cite{MaterialsProject}, or crystallographic databases such as ICSD~\cite{ICSD}, COD~\cite{COD1, COD2}, MPDS~\cite{MPDS1, MPDS2, MPDS}. While the user is responsible for providing these fundamental structures, the workflow progresses with automatic data augmentation to enhance dataset diversity without requiring exhaustive manual curation.

\subsubsection{Dataset augmentation \label{sec:augmentation}}
In the dataset augmentation stage, additional structures are generated by manipulating the initial set $\left\{\mathcal{X}\right\}^{(0)}$.
All manipulations can be controlled through customizable parameters to tailor the augmentation process according to specific user needs; we group them in the following categories:

\begin{itemize}
  \item {\em Supercells}: Initial structures are replicated aiming to ensure cells larger than a minimum threshold value (default: 18 \AA, corresponding to twice the MACE default receptive field) while keeping the total atom count below a user-defined maximum limit (default: 450 atoms).
    
  \item {\em Random distortions}: Atomic positions are perturbed with random displacements, where the magnitude follows a uniform distribution up to a user-defined fraction (default: 30\%) of the original nearest-neighbor distance. This introduces configurations away from equilibrium while preventing unphysically close atoms, which could lead to large forces that are difficult—and potentially uninformative—for NNIPs to learn.

  \item {\em Strain}: Strain can be applied to crystal structures by rescaling lattice parameters  by a factor randomly sampled from uniform distribution between a user-defined range (default from -20\% to +60\%). This is key for predicting elastic properties.

  \item {\em Vacancies}: Vacancies are created by removing atoms at randomly selected sites, to learn about defect energetics and local relaxations around missing atoms. By default, vacancies are introduced in 30\% of the randomly distorted structures, with 2 atoms removed for each configuration.
    
  \item {\em Clusters}: Atomic clusters are constructed by assembling atoms with a next-neighbor distance between 1 and $\sqrt{3}$ times a user specified distance (default: 1.5 \AA). This creates non-periodic environments that can become useful for training potentials capable of describing isolated molecules or clusters, and more in general surface or edge terminations.
    
  \item {\em Slabs}: Slabs are created by cutting bulk supercells along selected crystallographic directions (default: (111), (110) and (100)) ensuring a minimum slab thickness (default: 10 \AA) unless a maximum number of atoms (default: 450) is reached. Those structures allow the NNIPs to learn and predict surface energetics, surface-specific forces, reconstructions and other relaxation phenomena.
    
  \item {\em Isolated atoms}: Single-atom configurations are included to establish reference energies for accurate calculations of the dissociation limit. These structures are computed using the same DFT settings as the rest of the dataset to ensure consistency, though we note that the procedure does not take into account the actual magnetic configuration for some atomic species.
  
  \item {\em Atomic substitutions}: For multi-species systems, randomly selected atoms of different elements are swapped to create chemical disorder and explore different local chemical environments. The user can define both the fraction of previously generated structures to undergo substitutions (default: 20\%) and the number of swaps per structure as a fraction of total atom count (default: 20\%). This helps to explore various chemical environments, substitutional defects, and atomic site preferences, ensuring robustness across different chemical compositions within the same structural motif.

  \item {\em Alloys}: Alloy configurations are generated by randomly mixing atomic species (specified via \texttt{alloy\_species}), while optionally keeping some species fixed (specified via \texttt{fixed\_species}). If no target compositions are specified, the workflow samples random alloy configurations with concentrations spanning the full range from 0 to 1.
\end{itemize}

The presented dataset augmentation techniques are applied by default; however, users can choose to apply only a subset of them or even skip the dataset augmentation stage altogether. 
After this stage the resulting dataset $\left\{\mathcal{X}\right\}^{(1)}$ integrates structures of different dimensions and boundary conditions: fully periodic (bulks), partially periodic (surfaces, nanowires, or 2D materials), and non-periodic (molecules, clusters) configurations. However, since DFT and MD calculations are performed in full periodic boundary conditions, for all structures that are non-periodic at least along one direction, the workflow ensures the presence of an appropriate vacuum buffer (default: 15 \AA\ thick) along such directions in order to eliminate spurious interactions between periodic images.

\subsubsection{\textit{Ab Initio} Labelling}

After the data augmentation stage, AiiDA-TrainsPot starts the active learning loop, which is represented by the orange circle in Fig.~\ref{Fig1}. 
Each structure $\mathcal{X}_i$ in the augmented dataset is labeled through DFT calculations to obtain high-fidelity reference values for energies, forces, and stress tensors. We use the compact notation $\mathcal{L}^{DFT}(\mathcal{X}_i) = 
\left( E(\mathcal{X}_i), \mathbf{F}(\mathcal{X}_i), \boldsymbol{\sigma}(\mathcal{X}_i)
\right)$ to represent these computed properties. In subsequent sections, we denote specific quantities of interest as $\mathcal{L}_\alpha$ where $\alpha \in \{E, \mathbf{F}, \boldsymbol{\sigma}\}$.
Ultimately, \emph{ab initio} calculations directly determines an upper bound for the accuracy and precision of the trained NNIPs. While the overall accuracy is typically limited by considerations of computational efficiency and resources, precision can be substantially improved by enforcing well converged calculations and a consistent choice of key simulation parameters over cells of different sizes and dimensions. In this context, even if the workflow allows users to have full control over the DFT level of theory, by default AiiDA-TrainsPot enforces the use of well established simulations protocols originally introduced for high-throughput calculations. In particular, PBE pseudopotentials and cutoff parameters are given by version 1.3 of the branch of the Standard Solid State Pseudopotentials (SSSP) library optimized for precision \cite{Prandini2018,paw,pslib1}, while the reciprocal space k-point density and smearing follow the \textit{stringent} protocol defined by Nascimento et al.~\cite{Nascimento2025}.

By default, the workflow does not incorporate additional van der Waals (vdW) corrections at the DFT level, since NNIPs would in principle require rather large radial cutoffs to accurately learn long-range interactions from the training data. However, for systems where dispersion forces are critical (e.g., layered materials or molecular crystals), users can enable empirical vdW corrections (such as Grimme-D2, D3) during subsequent MD simulations \cite{Drautz}.

\subsubsection{Training neural-network interatomic potentials}
The labeled dataset $\left\{\mathcal{X}_i,\mathcal{L}^{DFT}(\mathcal{X}_i)\right\}^{(1)}$ is used to train a committee of $M$ NNIPs $\{\Phi_j\}_{j=1}^M$, each with identical architecture but initialized with different random seeds. Prior to training, all structures are systematically partitioned into three subsets, ensuring representative sampling across different structural motifs while maintaining similar distributions of atomic environments: 
\begin{itemize}
\item[-] {Training set} (default: 80\%): Used for model parameter optimization through gradient-based learning; 
\item[-] {Validation set} (default: 10\%): Used for hyperparameter tuning, early stopping decisions, and selection of optimal checkpoints during training; 
\item[-] {Test set} (default: 10\%): Reserved exclusively for final model evaluation, providing an unbiased assessment of generalization performance.
\end{itemize}

Throughout the active learning iterations, each structure remains in its initially assigned set, ensuring that the test set remains completely independent from the training and validation sets for reliable performance assessment. The training is performed using either MACE or Metatrain~\cite{metatrain} with default hyperparameters, though these can be fully customized by the user of AiiDA-TrainsPot to suit specific requirements.

\subsubsection{Exploration by molecular dynamics}
After a committee of NNIPs is trained, the workflow employs MD simulations to systematically explore the potential energy landscape. This exploration phase is critical for identifying configurations where the NNIPs might have insufficient accuracy, thus guiding the selection of additional structures for \emph{ab initio} calculations in subsequent training iterations.

At each iteration, one NNIP ($\Phi_1$) is randomly selected from the committee to perform the MD simulations. By default, a set of 20 structures is randomly sampled from the initial augmented dataset $\{\mathcal{X}\}^{(1)}$ to serve as starting configurations. This selection strategy deliberately uses structures from the initial dataset rather than from the most recent active learning iteration dataset to ensure sampling of a configuration space that is not biased by previous explorations and remains representative of the user's target application domain. Users can customize this selection by specifying either the number of structures to sample or providing an explicit list of starting configurations.

All MD simulation parameters are fully customizable, including ensemble type, temperature, pressure, simulation time, and timestep. By default, simulations run for 10 ns with a 1 fs timestep, while temperatures and pressures range from 0 to 1000 K and -5 to 5 kbar respectively, systematically sampling diverse thermodynamic conditions. AiiDA-TrainsPot automatically selects an isothermal-isobaric (NPT) ensemble with barostats acting only on periodic directions for bulk systems, or an isochoric-isothermal (NVT) ensemble for non-periodic systems. Additionally, even though dispersion corrections are disabled by default, users can activate Grimme's D2 van der Waals correction~\cite{Grimme2006} during MD simulations, with coupling parameters automatically selected based on the atomic species present in the system.

To minimize correlations between sampled configurations, trajectory frames are extracted at regular intervals (default: 1 ns), ensuring the collection of statistically uncorrelated dataset that efficiently represent the accessible regions of the PES.

\subsubsection{Committee Evaluation \label{sec:evaluation}}
This stage aims at identifying structures that are poorly predicted by the NNIPs; those are good candidates to be labeled with \emph{ab initio} calculations and included in the training dataset. However, while Bayesian neural networks (NNs) come with a well-defined probabilistic uncertainty quantification, no such Bayesian error estimation can be defined for NNIPs~\cite{Abdar2021}. Here, similar to what has been done in recent works (e.g.,~\cite{Zhang2020, Behler2014, Chen2020}), we use the spread of predictions from a committee of NNIPs as a proxy for uncertainty quantification.
Hence, at this step, each structure $\mathcal{X}_k$ sampled from MD trajectories is evaluated by all NNIPs of the committee and we quantify the uncertainty of prediction through the committee disagreement metric $\mathcal{D}_\alpha({\mathcal{X}_k})$, which is calculated separately for each property $\alpha$ (energy, forces\textemdash averaged over all the atoms and components\textemdash and stress tensor):
\begin{equation}
  \mathcal{D}_\alpha({\mathcal{X}_k}) = \sqrt{\frac{1}{M}\sum_{j=1}^{M}\left(\mathcal{L}_\alpha^{\Phi_j}({\mathcal{X}_k}) - 
  \overline{{\mathcal{L}}}_\alpha^{\Phi}({\mathcal{X}_k})\right)^2}
\end{equation}

where $M$ is the committee size, $\mathcal{L}_\alpha^{\Phi_j}({\mathcal{X}_k})$ represents the prediction of property $\alpha$ from potential $\Phi_j$ for structure $\mathcal{X}_k$, and $\overline{{\mathcal{L}}}_\alpha^{\Phi}({\mathcal{X}_i})$ denotes the mean prediction across all committee members.
Structures exhibiting disagreement beyond a specific threshold $\tau^{thr}_\alpha$ are flagged as uncertain and prioritized for quantum mechanical labelling.

The error $\epsilon_\alpha({\mathcal{X}_k})$ with respect to the \emph{ab initio} labels is
\begin{equation}
  \epsilon_\alpha({\mathcal{X}_k}) = \sqrt{\frac{1}{M}\sum_{j=1}^{M}\left(\mathcal{L}_\alpha^{\Phi_j}({\mathcal{X}_k}) - \mathcal{L}_\alpha^{DFT}({\mathcal{X}_k})\right)^2}
\end{equation}
and is not, at least in principle, strictly correlated to the committee disagreement, although empirical evidence suggests they might be linearly related~\cite{Kahle2022} (as we show and discuss more in detail in Sec.~\ref{subsec:validation} through a validation study on carbon allotropes). After linear regression, the user-defined error tolerance threshold $\epsilon^{thr}_\alpha$ (default 1 meV/atom, 100 meV/\AA, 1 meV/\AA$^3$ for energy, forces and stress tensor respectively) is transformed into an equivalent disagreement threshold $\tau^{thr}_\alpha$ that serves as the selection criterion for new structures by $\tau^{thr}_\alpha = a_\alpha \epsilon^{thr}_\alpha$, where coefficients $a_\alpha$ are the slopes determined from fitting. The calibration, made at each iteration on the already \textit{ab initio} labeled structures and compared with $\mathcal{D}_\alpha({\mathcal{X}_i})$ computed with the last generation potentials, ensures that the uncertainty quantification mechanism remains effective throughout the iterative learning process, as the NNIP committee's overall accuracy improves with each active learning cycle. Notably, the calibrated committee disagreement $a_\alpha\mathcal{D}_\alpha({\mathcal{X}_k})$ can be used not only in the active learning scheme for the selection of worse predicted structure, but also in production runs, where it can provide an estimate of the uncertainty of the predictions of the NNIPs.

For large exploration sets $\{\mathcal{X}_k\}$, computational constraints often necessitate limiting the number of structures labeled in each active learning iteration. However, selecting only the structures with the highest disagreement values may not be optimal, as these could represent configurations far from the original dataset and the PES region of interest to the user. While such structures cannot be discarded a priori—since they may represent poorly predicted configurations that are nonetheless close to relevant regions of the PES\textemdash a more balanced selection strategy is needed. Therefore, we randomly select structures from those that exceed the threshold $\tau_\alpha$ for any property $\alpha$. This approach naturally prioritizes structures closer to the original dataset, as they are more likely to be represented in the configurations extracted from MD trajectories, while still capturing the most uncertain predictions.
By default, at each active learning iteration a maximum number of 1000 new structures are selected. The active learning loop continues until one of two termination criteria is met: either all structures exhibit disagreement below the specified threshold $\tau^{thr}_\alpha$, indicating the achievement of the desired confidence across the configuration space of interest, or the maximum number of iterations $L$ is reached. The final output of AiiDA-TrainsPot includes the latest NNIP committee $\left\{\Phi\right\}$, the full {\it ab initio} dataset $\left\{X,\mathcal{L}^{DFT}\right\}^{final}$ and quantitative RMSE performance metrics for each potential.

\subsection{Code implementation \label{sec:code}}

AiiDA-TrainsPot is built on the AiiDA infrastructure~\cite{AIIDA1, AIIDA2}, providing a robust framework for managing complex computational workflows with full tracking of data provenance. The platform automatically persists all computational steps, input parameters, and generated data, ensuring complete reproducibility of results. The framework efficiently orchestrates job submissions across high-performance computing resources while leveraging existing AiiDA plugins for quantum mechanical and classical simulations.

The workflow architecture follows a hierarchical design of nested AiiDA WorkChains (Fig.~\ref{Fig2}), enabling both end-to-end automation and selective execution of individual components. The top-level \texttt{TrainsPotWorkChain} coordinates five specialized sub-processes that correspond to the major stages of a training campaign:

\begin{itemize}
  \item \texttt{DatasetAugmentationWorkChain} for generation of highly-diverse training structures
  \item \texttt{AbInitioLabellingWorkChain} for quantum mechanical calculations
  \item \texttt{TrainingWorkChain} for training NNIP committees
  \item \texttt{ExplorationWorkChain} for exploration of the PES based on MD
  \item \texttt{EvaluationCalculation} for committee-based error estimation and identification of structures for which the NNIP yields predictions with low accuracy.
\end{itemize}

\begin{figure}[tbp]
  \centering

   \includegraphics[trim={0 0 0 0px},clip,width=0.49\textwidth]{workcflow_small_META.pdf}
  
  \caption{\label{Fig2}{\bf Schematic representation of the \texttt{TrainsPotWorkChain} with its various computational tasks.} The workflow begins with the initialization phase, where input structures, parameters, and control flags are set to determine which steps of the workflow 
  will be executed. If dataset augmentation is enabled, the \texttt{DatasetAugmentationWorkChain} generates additional configurations. Next, the workflow enters the active learning loop.
  Within each iteration, the \texttt{AbInitioLabellingWorkChain} is called to label newly generated configurations by performing automated electronic structure calculations (\texttt{PwBaseWorkChain}). The \texttt{TrainingWorkChain} is then invoked to train interatomic potentials using MACE (\texttt{MaceTrainWorkChain}) or Metatrain (\texttt{MetaTrainWorkChain}). Subsequently, the \texttt{ExplorationWorkChain} executes MD simulations via LAMMPS
  (\texttt{LammpsBaseWorkChain})
  to generate new configurations for further refinement. 
  The \texttt{EvaluationCalculation} assesses the performance of the trained models using calibrated committee disagreement to determine whether additional iterations are required or not.}
\end{figure}

A key advantage of our implementation strategy and of the use of AiiDA as a workflow engine is the extensive reuse of existing AiiDA plugins, minimizing duplication of software and ensuring robustness through well-tested components. The \texttt{AbInitioLabellingWorkChain} calls \texttt{PwBaseWorkChain} from the AiiDA-Quantum ESPRESSO plugin~\cite{AIIDA1, AIIDA2}, while the \texttt{ExplorationWorkChain} leverages \texttt{LammpsBaseWorkChain} from the AiiDA-LAMMPS plugin~\cite{aiidalammps}. For specialized functionality not available in existing plugins, we develop custom components such as the \texttt{DatasetAugmentationWorkChain}, which implements various structure manipulation techniques through dedicated \texttt{calcfunctions} based on the Atomic Simulation Environment (ASE)~\cite{ase-paper}. Similarly, the \texttt{TrainingWorkChain} interfaces with NN and ML training codes and can be configured by the user to choose the training instrument, either \texttt{MaceTrainWorkChain} for MACE or \texttt{MetatrainWorkChain} for Metatrain, handling all preprocessing, training, and postprocessing steps in a fully automated manner. We note that Metatrain currently implements both the PET architecture and other ML architectures not all based on neural-networks, such as Sparse Gaussian Approximation Potentials (GAP)~\cite{bartok_prl_2010} and Behler-Parrinello neural networks with SOAP features (SOAP BPNN)~\cite{bp_prl_2007}.

To optimize computational efficiency, AiiDA-TrainsPot implements parallel execution strategies within three main WorkChains (\texttt{AbInitioLabellingWorkChain}, \texttt{TrainingWorkChain}, and \texttt{ExplorationWorkChain}). Multiple DFT calculations, neural-network training sessions, and MD simulations are submitted concurrently, with results collected and analyzed collectively before proceeding to the next workflow stage.

\begin{figure}[tbp]
  \centering

   \includegraphics[trim={0 0 0 0px},clip,width=0.5\textwidth]{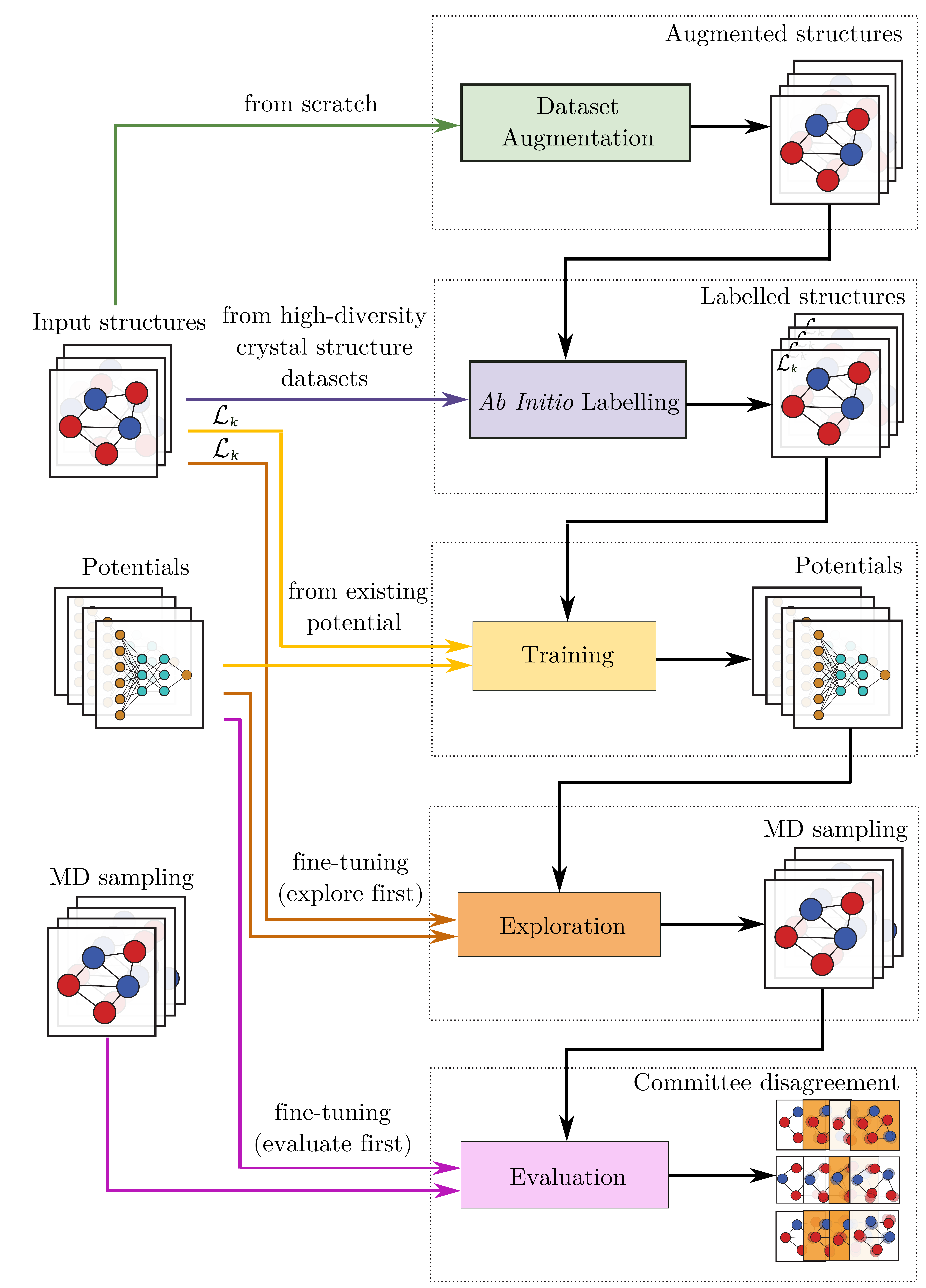}
\caption{\label{Fig3}{\bf Schematic representation of AiiDA-TrainsPot for different use cases}. Users can execute individual subtasks or the entire workflow depending on their specific requirements. The workflow can start from Dataset Augmentation to expand data diversity, \emph{Ab Initio} Labelling to perform DFT-based calculations of energy, force, and stress tensor, or Training to generate or improve existing machine-learned interatomic potentials. Users may also initiate Exploration for MD simulations or Evaluation to assess accuracy and performance.}

\end{figure}

This modular design provides several advantages beyond efficient execution. First, it offers maximum flexibility through multiple entry points, allowing users to bypass specific stages depending on their needs (see Fig.~\ref{Fig3}). For instance, users who would like to leverage access to existing datasets of labeled structures can proceed directly to training. The workflow also supports fine-tuning pretrained models, including foundation models.

Second, the architecture enables selective execution of individual components, permitting users to run specific stages independently. This is particularly valuable for scenarios such as structure labelling with DFT calculations without proceeding to the training phase, or for evaluating the accuracy of existing potentials on new trajectories using the committee disagreement.

Finally, the architecture maintains extensibility through AiiDA's plugin system, facilitating future integration with additional quantum engines, classical MD codes, or emerging ML frameworks. Individual components like \texttt{MaceTrainCalculation}, \texttt{MetaTrainCalculation} or \texttt{EvaluationCalculation} can be used as standalone tools outside the high-level WorkChain, making AiiDA-TrainsPot both a comprehensive platform and a flexible toolkit for specialized tasks in interatomic potential development.

To efficiently manage large datasets within the AiiDA framework, we introduce a custom AiiDA data type, \texttt{PESData}, a subclass of \texttt{aiida.orm.Data}. This specialized data type enables the storage and manipulation of extensive sets of atomic structures along with their associated properties, while maintaining full compatibility with AiiDA's provenance tracking system.

\texttt{PESData} offers significant advantages over existing data types such as \texttt{TrajectoryData}, which is limited to structures containing identical numbers of atoms. Our implementation can seamlessly handle datasets with varying numbers of atoms per structure, making it ideal for diverse training sets that include bulk materials, surfaces, clusters, and defect structures. Beyond structural information and labeled properties, \texttt{PESData} can also store custom metadata including references to the original data sources, committee evaluation results, accuracy metrics, computational parameters used for labelling, and other application-specific information.

For performance optimization, the dataset is stored in the AiiDA repository as an HDF5 file using the h5py library~\cite{hdf5}. The class implements Python iterators to read data in chunks, enabling efficient handling of large datasets without overwhelming memory resources. This approach is crucial when working with training sets containing thousands of structures with hundreds of atoms each, and to maintain high performance and usability throughout the NNIP development workflow.

\subsection{Validation \label{subsec:validation}}
To validate our workflow and demonstrate its capabilities, we consider two representative case studies. First, we carry out a fully automated training campaign of a NNIP for a comprehensive set of carbon allotropes, showcasing the ability of the workflow to handle materials with widely different bonding topologies and dimensionalities. Second, we investigate 2D transition-metal dichalcogenides, focusing on monolayer 
$\mathrm{W_xMo_{1-x}Te_2}$  alloys, to illustrate the flexibility of the approach for multicomponent systems and its accuracy in capturing structural phase transitions as a function of composition. 

\subsubsection*{Carbon allotropes}
As a first validation study, we showcase AiiDA-TrainsPot with an active learning training of a NNIP for all carbon allotropes. Carbon represents an ideal test case due to its rich structural diversity, which includes 0D fullerenes, 1D nanotubes, 2D graphene, various 3D crystals, including diamond and layered forms such as graphite. 

We conduct two independent validation runs, both initiated from the same set of 48 carbon structures primarily sourced from the Materials Cloud 2D and 3D databases~\cite{MaterialsCloud} (44 structures), supplemented with 4 selected low-dimensional structures (nanoribbons and fullerenes) from the dataset developed by Drautz \textit{et al.}~\cite{Drautz}.

The first run, designated as \textit{run A} or ``fast exploration'', is designed to assess the workflow's capability to explore the PES and study the convergence of the active learning scheme for increasingly large datasets. The run starts by augmenting structures near equilibrium configurations (see Methods Sec.~\ref{sec:Methods} for more details), that is followed by MD with a wide range of temperatures from 0 up to 5000 K. This also tests the model's ability to capture extreme conditions such as melting and formation of amorphous phases. As we want to explore convergence for rather large training datasets, which involve up to $10^4$ single-point DFT calculations, we employ here the AiiDA-QuantumESPRESSO \textit{fast} protocol~\cite{Nascimento2025} for DFT calculations~\cite{AIIDA1,AIIDA2}.

The second run, designated as \textit{run B} or ``accuracy and data-efficiency'', focuses, instead, on achieving rather accurate predictions for equilibrium and near-equilibrium conditions with small training datasets (around $2\times10^3$ calculations). The settings of data augmentation and the MD runs are refined for more efficient exploration, including smaller distortions of atomic positions, lower temperatures (up to 1000 K) and larger range of pressures (see Methods Sec.~\ref{sec:Methods}). As we aim at training an accurate NNIP in an efficient and automated manner, this run employs the default AiiDA-QuantumESPRESSO \textit{stringent} protocol: not only this ensures more precise DFT reference data, but also reduces  noise across the training dataset.

\begin{figure*}[tbp]
  \centering
    \includegraphics[width=1.0\textwidth]{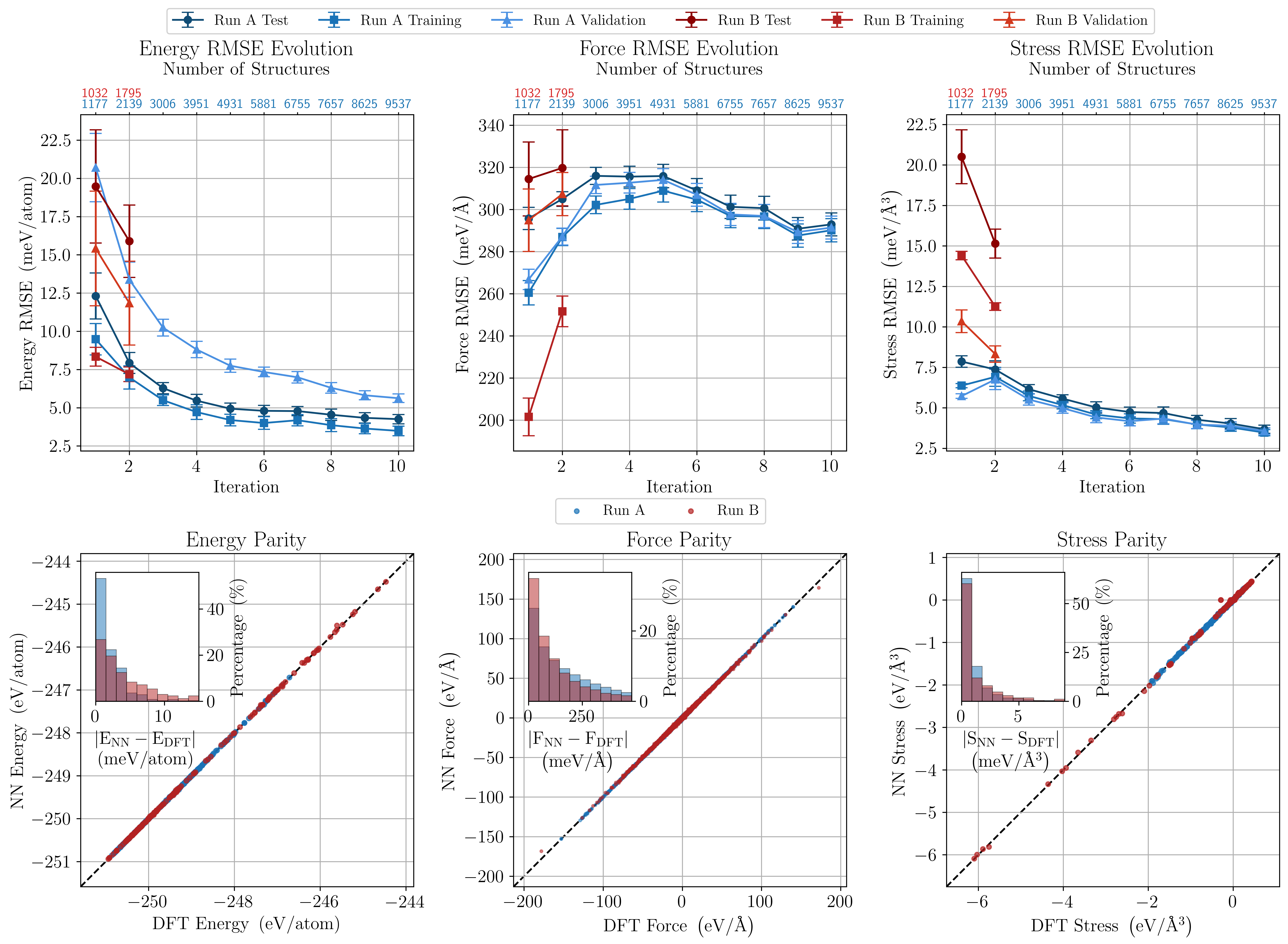}
\caption{\label{Fig4} {\bf Evolution of model accuracy over active learning.} Top panels: RMSE for energies, forces, and stress tensor components are shown across training, validation, and test sets. Bottom panels: Parity plots comparing NNIP predictions (latest iteration of active learning for both runs) to DFT reference values on the final test set. Cold and warm colors identify the results of run A (``fast exploration'') and run B (``accuracy and data-efficiency''), respectively. Error bars in the top panels represent the standard deviation across the model committee. Insets in the bottom panels show the error distribution histograms.}
\end{figure*}

\begin{figure*}[tbp]
  \centering
    \includegraphics[height=0.45\textwidth]{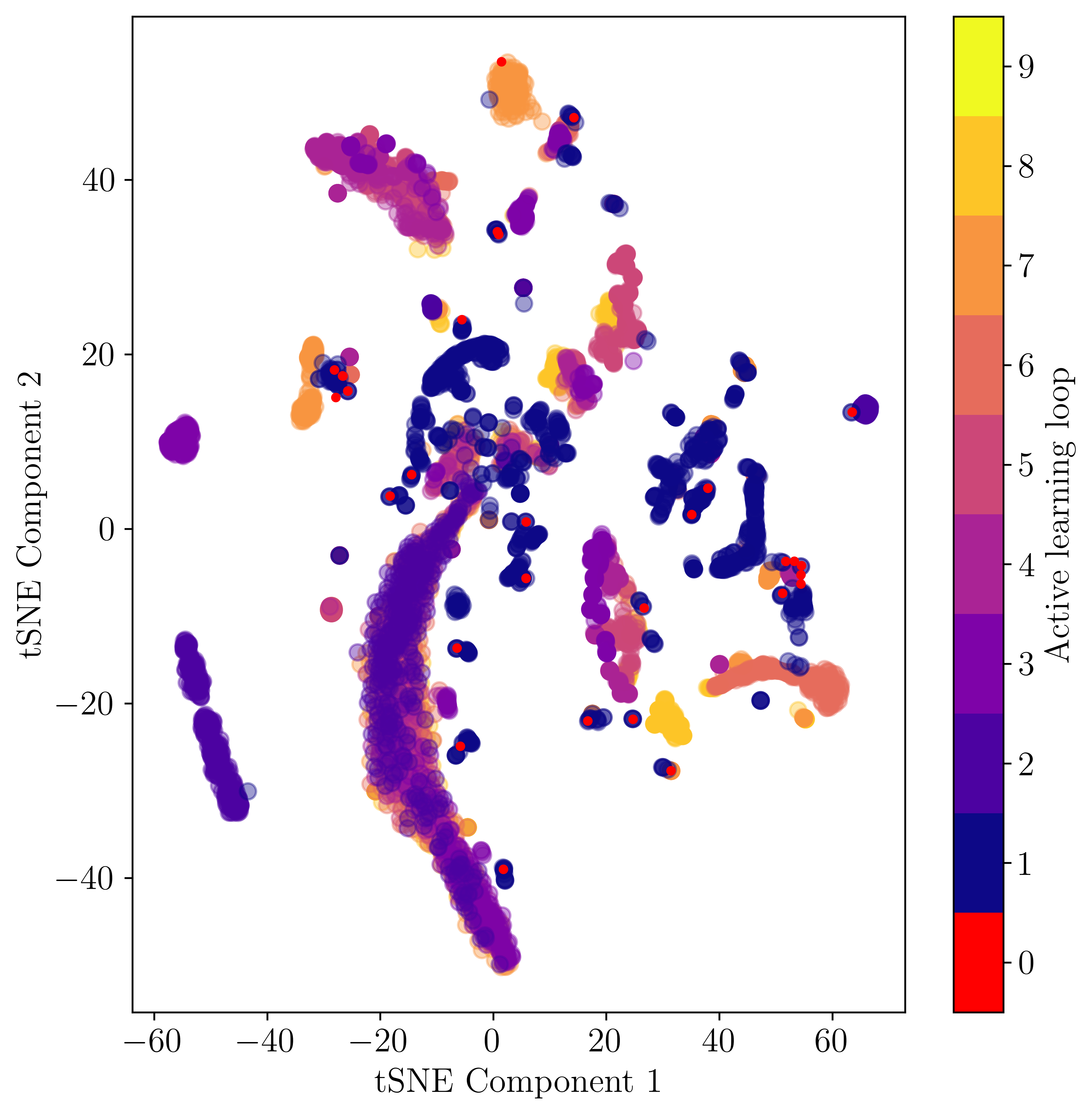}
    \includegraphics[height=0.45\textwidth]{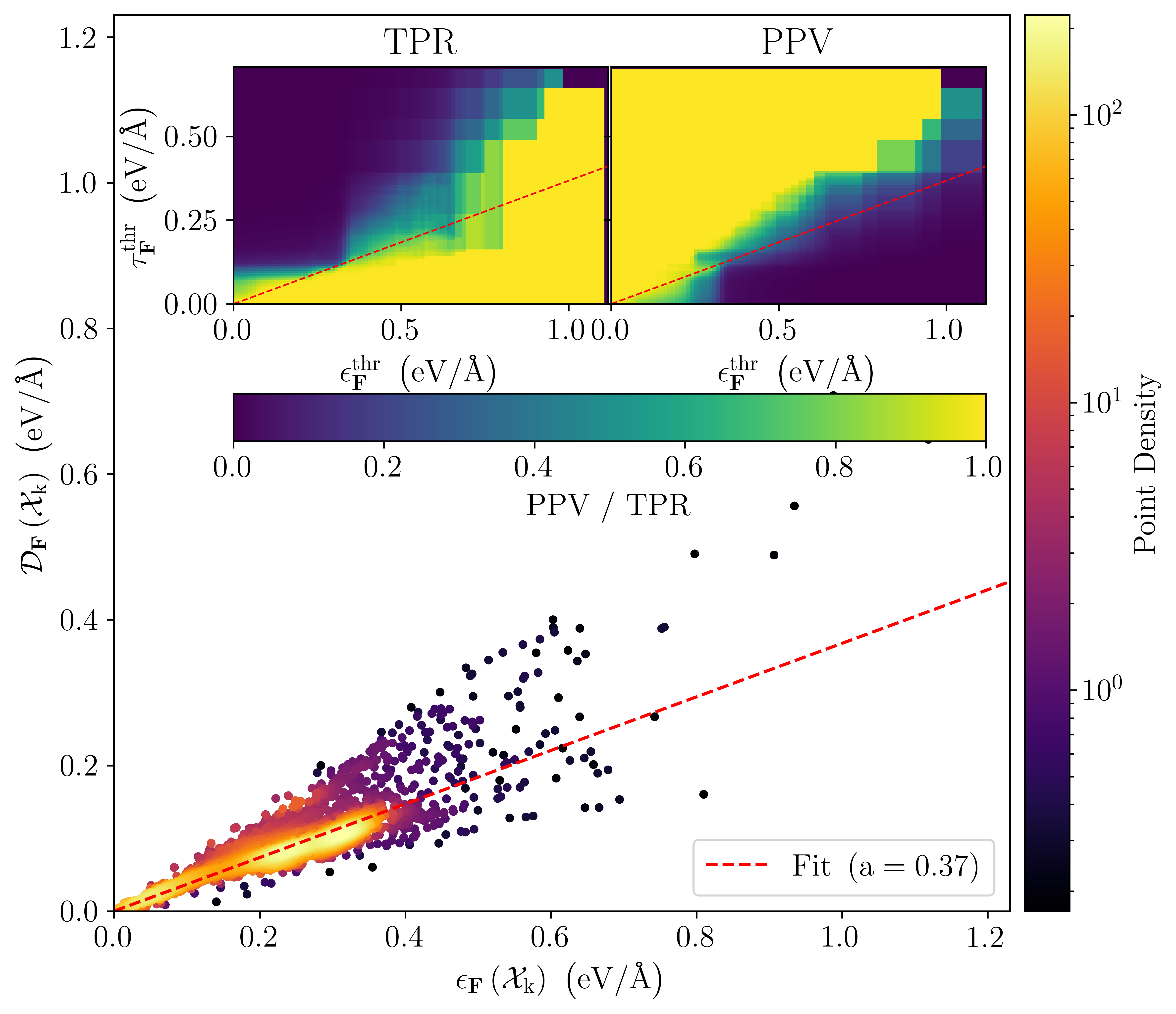}
\caption{\label{Fig5}{\bf Exploration of the potential energy surface (PES) and uncertainty quantification} (run A).
Left panel: t-SNE visualization of SOREP electronic-structure descriptors colored by active learning iteration, showing how the workflow systematically explores the PES through in two ways: by improving the sampling of known regions and by simultaneously exploring previously uncharted areas.
Right panel: Committee disagreement versus true error (deviation from DFT) across different active learning iterations, showing strong correlation between model uncertainty and actual prediction uncertainty. We calibrate committee disagreement with linear regression against true errors, which enables quantitative uncertainty estimation in large-scale applications where reference DFT calculations are not feasible. Insets show the True Positive Rate (TPR) and the Positive Predictive Value (PPV) as functions of disagreement and true error thresholds. Both metrics approach unity along the fitted correlation line (dashed red), i.e., they are simultaneously maximized by a structure selection strategy based on calibrated committed disagreement.}
\end{figure*}

The fast-exploration campaign runs for 10 iterations of active learning, Fig.~\ref{Fig4} reports the evolution of the model performance in cold colors. The top panels track the RMSE for energy, forces, and stress tensor components across training, validation and test sets, as a function of the active learning step and, in parallel, the dataset size that range from the initial 1,177 structures (iteration 1) to 9,537 structures in the final iteration. For energy and stress tensor components, we observe a consistent decrease in prediction errors as the active learning progresses. Interestingly, errors on forces increase from the first to the second iteration, and then decrease monotonically for all the following iterations, suggesting that the active learning strategy first explores novel regions of the PES that require more data to learn. This is supported by the data analysis based on dimensionality reduction  discussed later, which shows that early iterations sample an increasing number of distinct structural prototypes. The final RMSE values reach 4.3 meV/atom for energies, 293.0 meV/\AA\ for forces, and 3.7 meV/\AA$^3$ for stress tensor components on the test set, comparable to those reported in Ref.~\cite{Drautz} for a ML potential trained and tested on all carbon allotropes. The error bars in Fig.~\ref{Fig4} represent the standard deviation across the model committee, which also decrease with iterations as the model becomes more consistent and robust. The bottom panels display parity plots comparing DFT reference values against NNIP predictions for the test set after the final iteration. The tight clustering of points along the diagonal line, together with the error distribution histograms (insets), demonstrates excellent agreement between the ML predictions and DFT calculations across all evaluated properties.

To better understand how our active learning strategy explores the potential energy landscape, we analyze data diversity in the training dataset by using the kinetic Spectral Operator Representation (SOREP) descriptor~\cite{Zadoks2024}. The kinetic SOREP provides a compact electronic-structure fingerprint based on the density of states computed for the kinetic energy operator, which is evaluated on a basis set made of a customized version of the atomic natural orbitals (ANO) in terms of contracted Gaussian-type orbitals (cGTO)~\cite{pritchard_new_2019, feller_role_1996,schuchardt_basis_2007, roos_main_2004,roos_tm_2005,roos_act_2005,roos_lanth_2008}. We then use t-SNE (t-distributed Stochastic Neighbor Embedding) to visualize the high-dimensional SOREP descriptors in 2D space, with points colored according to the active learning iteration in which they were generated (Fig.~\ref{Fig5}a). The clustering pattern reveals that early iterations (iterations 1-2) sample distinctly different and broad regions of the configuration space compared to the initial dataset (depicted as iteration 0 in Fig.~\ref{Fig5}a). This explains the initial increase in force errors, as the model encounters novel atomic environments with rather different electronic structures, requiring additional data to be learned accurately. In later iterations, after thorough sampling of these initial regions, the workflow begins exploring new domains. While the decrease in errors after the initial iterations can be attributed to good sampling of primary regions, the reduction is modest as exploration of new regions continues concurrently. Although the final model achieves rather good accuracy, further active learning iterations could be performed to explore a wider region of the PES and enhance even further model performance.

Beyond understanding the exploration strategy through SOREP analysis, we evaluate the effectiveness of using committee disagreement as an uncertainty metric for steering the growth of the training dataset. Figure \ref{Fig5}b demonstrates the correlation between committee disagreement $\mathcal{D}_\alpha(\mathcal{X}_k)$ and true prediction errors $\epsilon_\alpha(\mathcal{X}_k)$ at the final active learning iteration. The analysis reveals an approximately linear relationship between these quantities, confirming that committee disagreement can serve as a reliable proxy for accuracy~\cite{Kahle2022}, although the proportionality constant is far from being unity. As anticipated in Sec. \ref{sec:evaluation}, we introduce a calibration factor $a_\alpha$ determined by linear regression, which transforms user-defined error tolerances $\epsilon^{thr}_\alpha$ into equivalent disagreement thresholds $\tau^{thr}_\alpha$; the calibrated committee disagreement is then used to select the structures to calculate from first principles.
This calibration procedure ensures that the committee disagreement threshold appropriately reflects the true prediction errors, addressing cases where the uncalibrated disagreement metric might either overestimate (as observed for forces in Fig.~\ref{Fig5}b) or underestimate (as seen for energies and stress in our validation test) the actual deviations from DFT.

To quantitatively assess the reliability of this approach, we analyze two key metrics shown in the insets of Fig.~\ref{Fig5}b: the True Positive Rate (TPR) and Positive Predictive Value (PPV) as a function of the disagreement and the true error thresholds. The TPR is defined as:
\begin{equation}
\mathrm{TPR} = \frac{\mathrm{TP}}{\mathrm{TP} + \mathrm{FN}}
\end{equation}
where $\mathrm{TP}$ is the number of true positives ($\mathcal{D}_\alpha(\mathcal{X}_k) > \tau^{thr}_\alpha$ and $\epsilon_\alpha(\mathcal{X}_k) > \epsilon^{thr}_\alpha$) and $\mathrm{FN}$ is the number of false negatives ($\mathcal{D}_\alpha(\mathcal{X}_k) < \tau^{thr}_\alpha$ and $\epsilon_\alpha(\mathcal{X}_k) > \epsilon^{thr}_\alpha$). The PPV is defined as:
\begin{equation}
\mathrm{PPV} = \frac{\mathrm{TP}}{\mathrm{TP} + \mathrm{FP}}
\end{equation}
where $\mathrm{FP}$ is the number of false positives ($\mathcal{D}_\alpha(\mathcal{X}_k) > \tau^{thr}_\alpha$ and $\epsilon_\alpha(\mathcal{X}_k) < \epsilon^{thr}_\alpha$). These metrics quantify the reliability of using committee disagreement for structure selection. A high TPR indicates the approach successfully identifies structures with large true errors, while a high PPV confirms that selected structures genuinely require additional training. The analysis shows that along the fitted correlation line (dashed red line), i.e., using a calibrated committee disagreement, both TPR and PPV remain close to unity, particularly for force predictions up to several hundred meV/\AA\textemdash~which is the typical range for accuracy thresholds and where most data points lie. Not only this shows that calibrated committee disagreement effectively identifies the structures requiring additional \emph{ab initio} calculations, but it shows to be an optimal strategy that simultaneously maximize both TPR and PPV, hence minimizing both the number of false positives and false negatives.

While the first validation study demonstrates the effectiveness of our active learning strategy, we observe diminishing returns in RMSE improvement as the number of iterations and the size of the dataset increase, due to the exploration of larger and larger regions of the PES. This suggests that strategic optimization of initial structures and data augmentation parameters can enhance the NNIP accuracy for the target application, while performing very few iterations of active learning. Therefore, for the second validation\textemdash focused on accuracy and data efficiency\textemdash we target the description of polymorphs near equilibrium with a refined data augmentation strategy combined with constrained temperature ranges in MD simulations: strain range is increased with respect to the previous run, while atomic rattling is reduced and MD temperatures range up to 1000 K (see Methods~\ref{sec:Methods} for details). This approach delivers high-quality potentials in just two active learning iterations and about 1800 training configurations, hence making more affordable and sustainable the use of the \textit{stringent} protocol for DFT calculations, which is computationally more expensive.

The accuracy metrics are reported as warm colors in Fig.~\ref{Fig4}: the NNIPs score (test set RMSE) 15.9 meV/atom for energies, 319.8 meV/\AA\ for forces, and 15.1 meV/\AA$^3$ for stress tensor components. While the overall errors on the test set seem comparable to the first run, we target here accurate energetic and vibrational properties, which are shown later to be in good agreement with DFT.

\begin{figure}[tbp]
    \centering
        \includegraphics[width=0.5\textwidth]{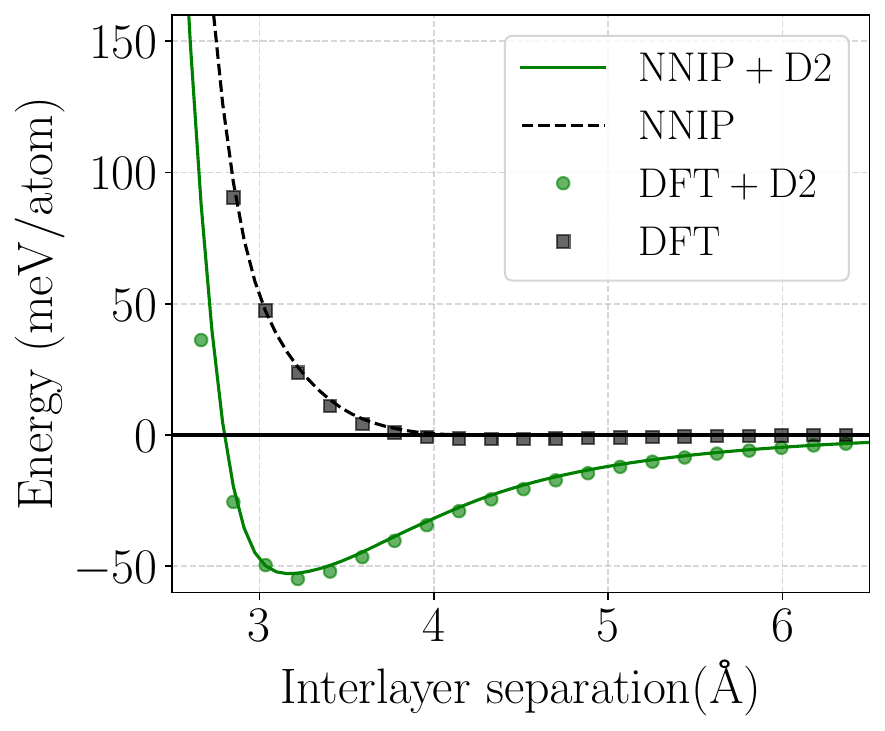}
  
\caption{\label{Fig6}Energy as a function of interlayer separation in graphite. The NNIP-predicted binding curve closely follows the DFT reference, both with and without Grimme's D2 dispersion corrections. Decoupling van der Waals interactions from the NN  helps to keep the descriptor more local and reduce the computational cost of training and using NNIPs.}
\end{figure}

Although our training data is obtained with the semi-local Perdew-Burke-Ernzerhof (PBE) functional~\cite{Perdew_PRL_1996}, we efficiently include long-range van der Waals interactions by adding Grimme's D2 dispersion corrections~\cite{Grimme2006} on top of the NNIPs. We check that the approach works in practice by calculating the energy profile as a function of interlayer distance in graphite (see Fig.~\ref{Fig6}), comparing the NNIP with PBE\textemdash both with and without D2 corrections, where the in-plane lattice parameters are fixed with DFT structural optimization. Except for defect formation energies, the following validation tests are all performed with Grimme's D2 dispersion correction applied on top of the NNIP and compared with DFT calculations that include D2 corrections as implemented in \textsc{Quantum ESPRESSO}.

\begin{figure*}[tbp]
  \centering
   \includegraphics[width=1\textwidth]{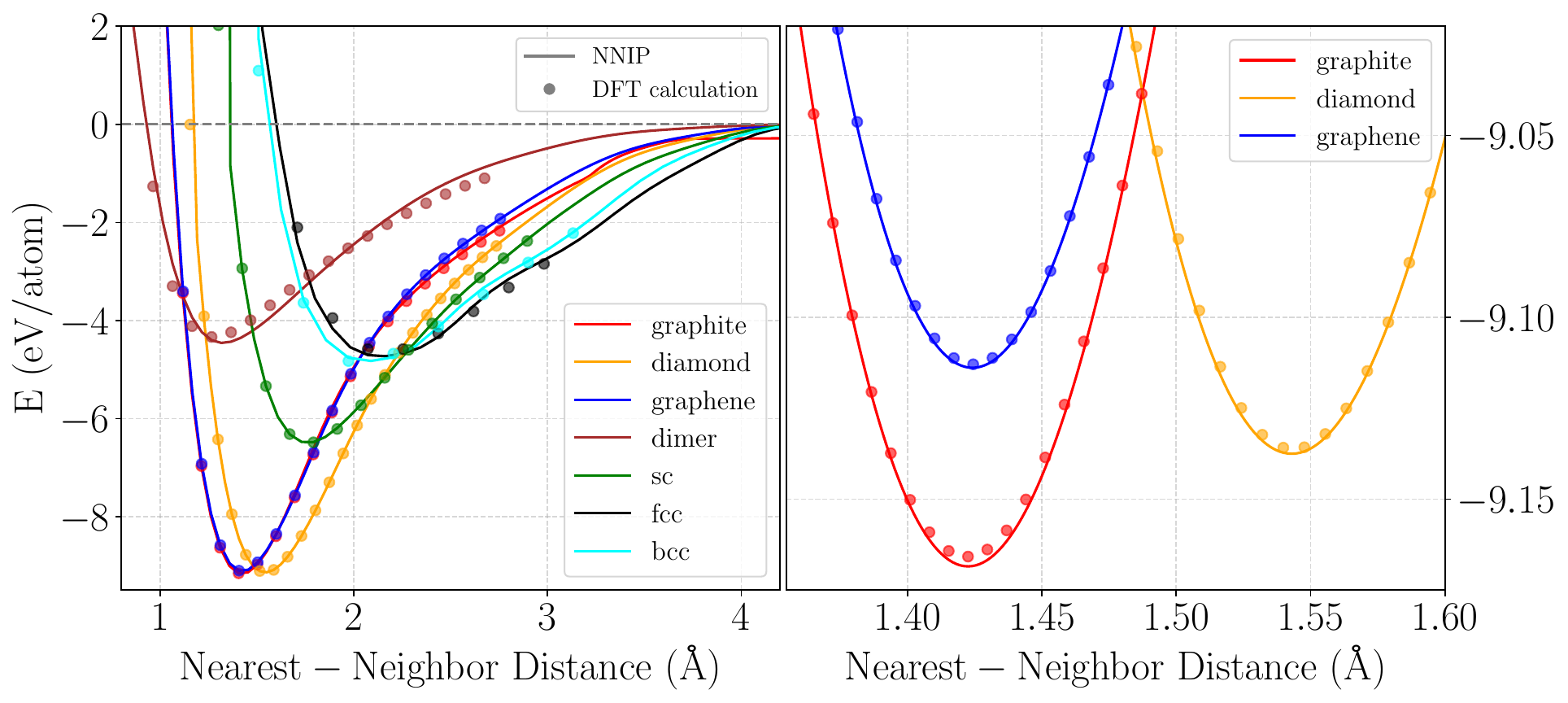}
\caption{\label{Fig7} Equations of state. Left panel: Binding energy per atom as a function of nearest-neighbor bond distance for various carbon structures, comparing predictions based on the NNIP with DFT calculations. Right panel: Zoomed-in view for graphite, graphene, and diamond; the model preserves the correct stability ordering at the meV level.}
\end{figure*}

Figure~\ref{Fig7} shows the equation of states for various carbon allotropes, including graphite, graphene, diamond, dimer, simple cubic (sc), face-centered cubic (fcc), and body-centered cubic (bcc) structures. Binding energies are evaluated with reference to the energy of isolated atoms. This benchmarking approach is widely adopted in the literature~\cite{Drautz, Ko2021}, as it offers a compact yet informative way to assess how accurately a potential reproduces bonding behavior across diverse local geometries.

The left panel of Fig.~\ref{Fig7} shows the excellent agreement between NNIP predictions and DFT references in around equilibrium; discrepancies become more noticeable at extreme bond compressions or expansions, which correspond to configurations that are underrepresented in the training data. Notably, the EOS for the dimer is reasonably accurate, even if no dimer configurations were included in the initial training set. The right-hand panel compares the EOS for graphite, graphene, and diamond, focusing on their relative energetic ordering. This inset is especially informative because graphite and diamond exhibit nearly degenerate formation energies in DFT. The NNIP correctly reproduces the energy hierarchy, demonstrating its ability to capture subtle thermodynamic trends. Similar benchmarking practices have been applied in the development of the ACE interatomic potentials \cite{Drautz}.

\begin{figure}[tbp]
    \centering
        \includegraphics[width=0.5\textwidth]{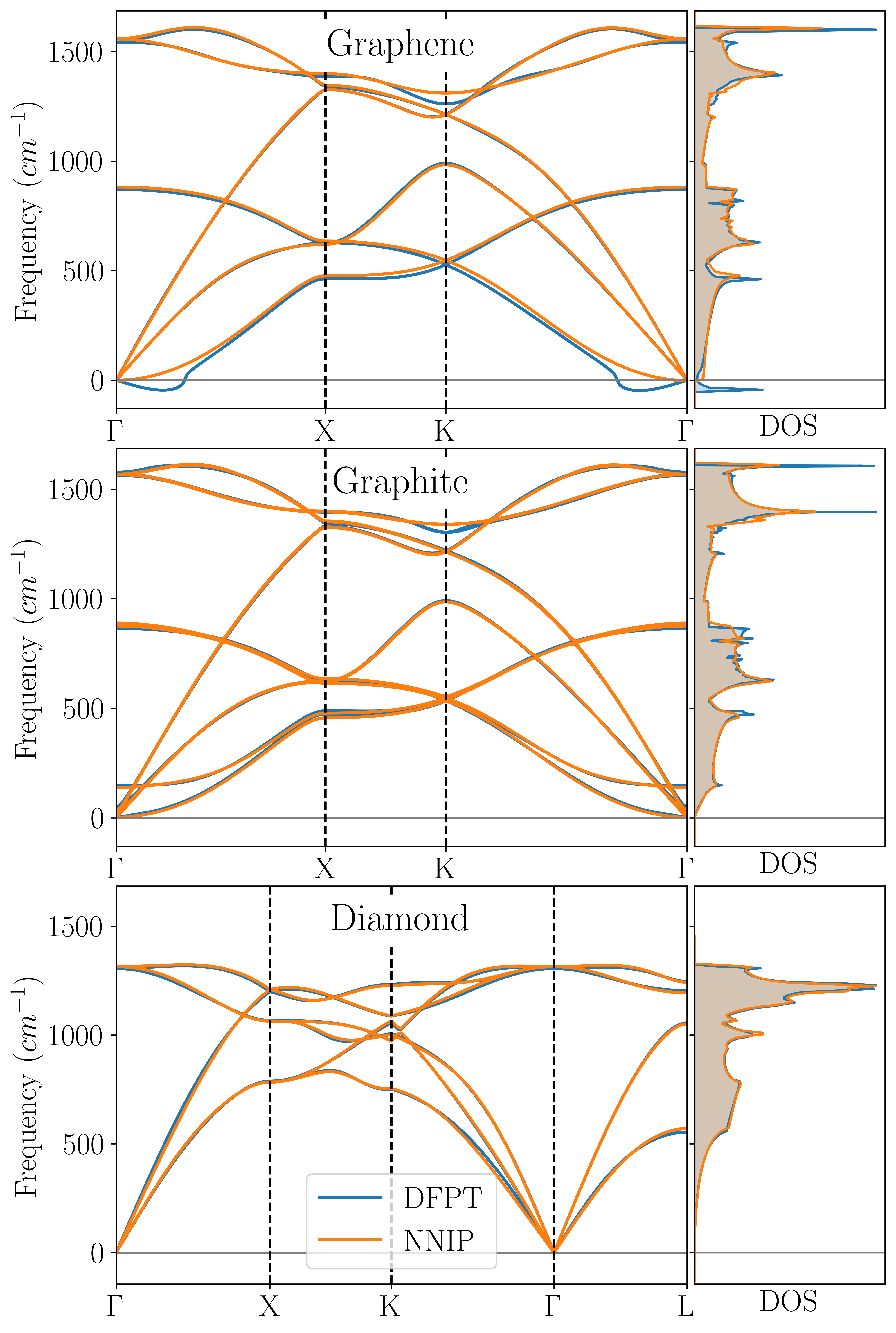}
\caption{\label{Fig8}Phonon dispersion and density of states for graphene, graphite, and diamond. The comparison between NNIP predictions and DFPT calculations shows good agreement for all three carbon allotropes.}
\end{figure}

\begin{table}[h!]
\centering
\begin{tabular}{lccc}
\hline
\textbf{Relax-Energy} & \textbf{Monovacancy} & \textbf{Divacancy} & \textbf{Stone–Wales} \\
\hline
NNIP-NNIP & 7.30 & 7.91 & 4.74 \\
NNIP-DFT  & 8.03 & 7.47 & 4.67 \\
DFT-NNIP & 7.72 & 8.00 & 4.74 \\
DFT-DFT  & 7.72 & 7.39 & 4.64 \\
\hline
\end{tabular}
\caption{\label{Table1}Formation energies (eV) of representative defects in graphene: monovacancy, divacancy, and Stone–Wales. 
The first two rows correspond to structures relaxed with the NNIP, with subsequent energy evaluation using either the NNIP (row 1) or DFT (row 2). 
The third and fourth rows correspond to structures relaxed with DFT, with subsequent energy evaluation using the NNIP (row 3) or DFT (row 4).}
\end{table}

To assess the transferability of our NNIP to defective structures, we compute the formation energies of three representative point defects in graphene: the monovacancy, divacancy, and Stone-Wales defect. The formation energy is defined as:
\[
E_{\text{form}} = E_{\text{defected}} - E_{\text{pristine}} + n \cdot E^0,
\]
where $E_{\text{defected}}$ and $E_{\text{pristine}}$ are the total energies of the relaxed defective and pristine graphene supercells, respectively; $n$ is the number of carbon atoms removed ($n = 0$ for Stone–Wales); and $E^0$ is the chemical potential of a carbon atom, estimated from bulk graphene.

As summarized in Table~\ref{Table1}, we evaluate defect formation energies using four complementary approaches: structures relaxed with the NNIP and subsequently evaluated with either the NNIP or DFT, and structures relaxed with DFT and evaluated with either the NNIP or DFT. This protocol allows us to separately assess the contributions of structural relaxation versus energy evaluation to the overall accuracy.

For all the cases considered, the calculated formation energies agree well with literature values~\cite{Drautz, Banhart2011} and our DFT references. For the divacancy and Stone-Wales defects, residual NNIP-DFT differences stem mainly from energy evaluation and not from the relaxation method. For the monovacancy, however, differences in the relaxed structures matter more: spin-unpolarized DFT yields a Jahn-Teller-like reconstruction with a slight out-of-plane displacement of one dangling atom. This subtle rearrangement is not fully reproduced by the NNIP, which returns indeed highly accurate energetics on the DFT-relaxed structure, but describes less accurately the local PES curvature. Instead, the NNIP predicts a completely flat configuration to be the energetically more stable for the monovacancy.
However, we should note that a physically accurate description of vacancies in graphene\textemdash particularly monovacancies\textemdash would anyway require spin-polarized DFT calculations to properly account for the unpaired $\pi$ orbitals~\cite{Palacios2008}, which were not used for our reference dataset.

We present in Fig.~\ref{Fig8} the phonon dispersion and density of states for graphene, graphite, and diamond. The comparison between NNIP predictions and density-functional perturbation theory (DFPT)~\cite{Baroni_RMP_2001} calculations demonstrates good agreement across all three carbon allotropes, confirming the potential's ability to accurately capture vibrational properties. The observed small imaginary ZA phonons near $\Gamma$ are a well-known numerical issue for 2D materials, which can be solved by adopting prohibitively tight parameters and in particular very high plane-wave cutoffs~\cite{Mounet2020}. Notably, despite being trained on DFT data exhibiting this, the NNIP does not display such unphysical behavior.

\begin{figure}[tbp]
    \centering
        \includegraphics[width=0.5\textwidth]{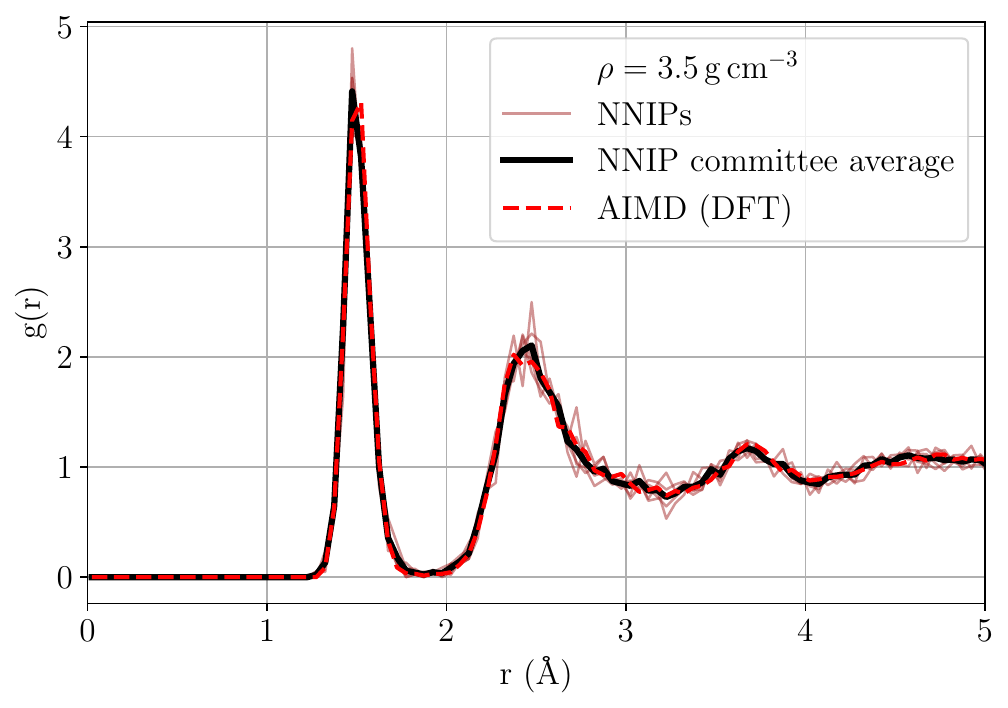}
  
\caption{\label{Fig9}Radial distribution functions $g(r)$ of amorphous carbon structures with a density of 3.5~g/cm$^3$, generated using a melt-and-quench protocol and analyzed with our five NNIPs (thin brown lines) and their committee average (thick black line). The results demonstrate that our potentials accurately reproduce the AIMD, including the positions of the first and second peaks, and the characteristic nonzero minimum at $\sim 1.9$~Å.}
\end{figure}

In order to further assess the transferability of our NNIP, we consider amorphous carbon, which features a disordered network of mixed sp$^2$ and sp$^3$ bonds that was not explicitly included in the rattled and strained crystalline configurations included in the training set. We compute the radial distribution function (RDF), $g(r)$, which characterizes short-range order in disordered systems, on independent amorphous configurations at a density of 3.5~g/cm$^3$ that have been obtained using the melt-and-quench procedure described in Ref.~\cite{Shaidu2021}.
Figure~\ref{Fig9} compares the RDFs for the committee of potentials and the committee average with reference {\it ab initio} molecular dynamics (AIMD) data from Ref.~\cite{amorphouscarbon}. 
The NNIPs closely reproduce the AIMD reference, accurately capturing the position and height of the first and second peaks, confirming the ability of our potentials to reliably describe the local structure of amorphous carbon.

\begin{figure*}[tbp]
  \centering
  \includegraphics[width=1\textwidth]{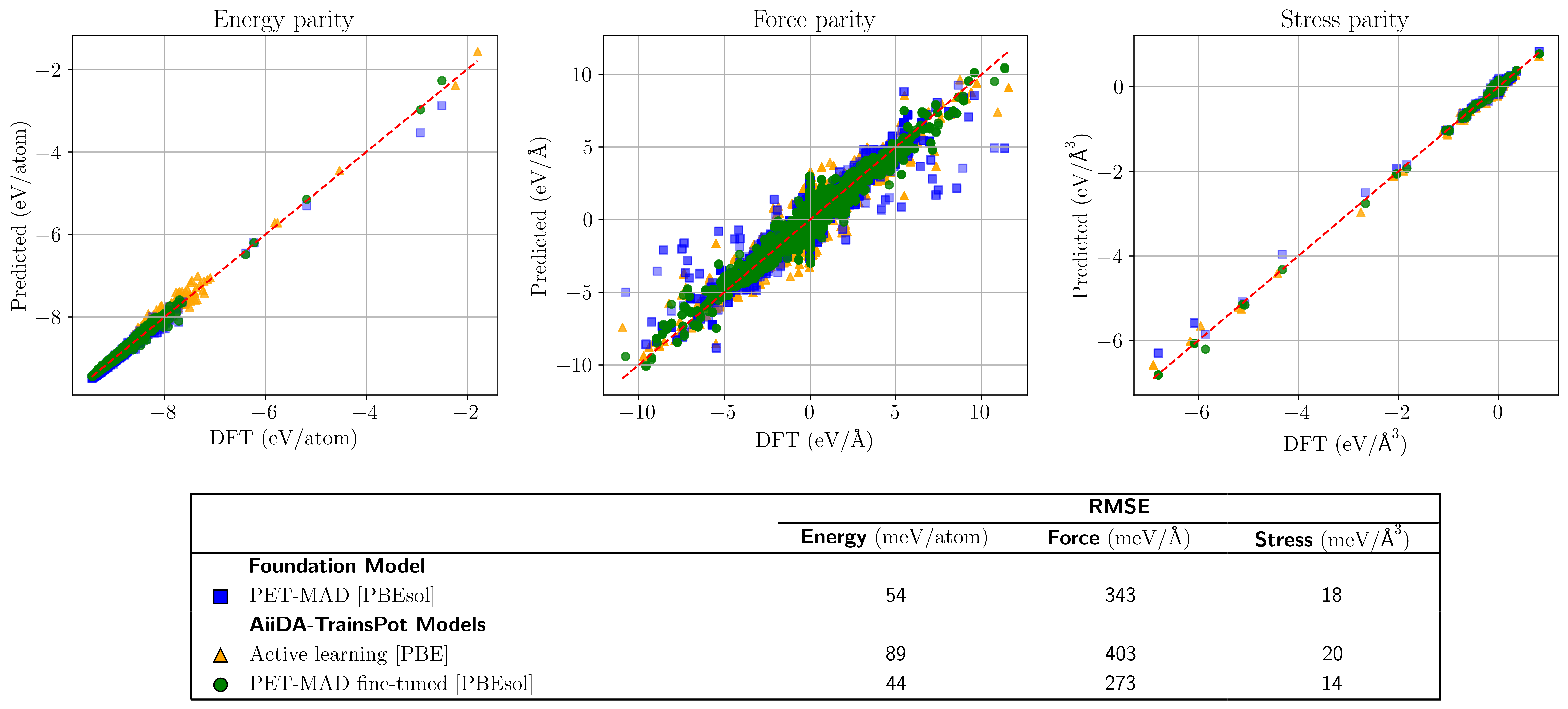}
  \caption{\label{Fig10}
  Parity plots for energy, forces, and stress on 98\% of the SACADA dataset, comparing three models:  
  (i) our MACE-based potentials generated in \textit{run B} (orange),  
  (ii) the universal PET-MAD potential (blue), and  
  (iii) the PET-MAD potential fine-tuned using the carbon dataset produced in \textit{run B} (green).  
  The dashed line indicates perfect agreement with the reference DFT data.  
  }
\end{figure*}

As a further transferability test, we evaluate one NNIP on the SACADA dataset, which contains 1635 carbon allotropes collected from the scientific literature~\cite{SACADA}. Out of the 1635, we exclude 13 structures due to unconverged DFT labeling. Then, we discard 19 structures where at least one of the potentials yields a committee-based disagreement larger than the following loose thresholds:
0.2 eV for energy, 50 eV/\AA{} for forces, and 1 eV/\AA$^3$ for the stress tensor. The remaining 1603 structures (98\% of the SACADA dataset) represent the overlap set between converged DFT simulations and successful NNIPs predictions, as required for a meaningful comparison. Here, we also evaluate the universal PET-MAD potential (v1.0.2)~\cite{petmad}, both in its original form and after fine-tuning on the carbon dataset generated in \textit{run B}. Since \textit{run B} was obtained with PBE functionals, while PET-MAD was trained with PBEsol~\cite{pbesol}, we relabel the SACADA dataset with PBEsol for evaluating the accuracy of PET-MAD potentials. In passing, we note how the modular structure of AiiDA-TrainsPot permits a seamless exchange of training engines (MACE or Metatrain) and architectures, the reuse of existing labeled datasets, and uniform evaluation procedures across models. 
We report the parity plots and RMSE in Fig.~\ref{Fig10}: the MACE potential trained from scratch with AiiDA-TrainsPot on 1795 structures performs similarly to the PET-MAD foundation model, with comparable error on forces and stress tensors while doing slightly worse on energies (89 vs. 54 meV). Notably, fine-tuning PET-MAD on the PBEsol-relabeled \textit{run B} datasets lowers the error on all metrics (energies, forcess and stress) both compared to the MACE potential trained from scratch and to PET-MAD.

\subsubsection*{Phase stability in monolayer $\mathrm{W_xMo_{1-x}Te_2}$ alloys}
As a further validation, we apply AiiDA-TrainsPot to a more challenging problem: the relative phase stability of $\mathrm{W_xMo_{1-x}Te_2}$  monolayer alloys. This system is well known to exhibit a composition-dependent transition between the H phase and $\mathrm{T'}$  phase, which can be driven by temperature, strain, doping, or alloying~\cite{Zhang2016}. In this benchmark, we specifically assess the ability of different models to reproduce the evolution of the formation-energy difference between the two phases as a function of alloy composition.

\begin{figure}[tbp]

  \centering
  \includegraphics[width=0.5\textwidth]{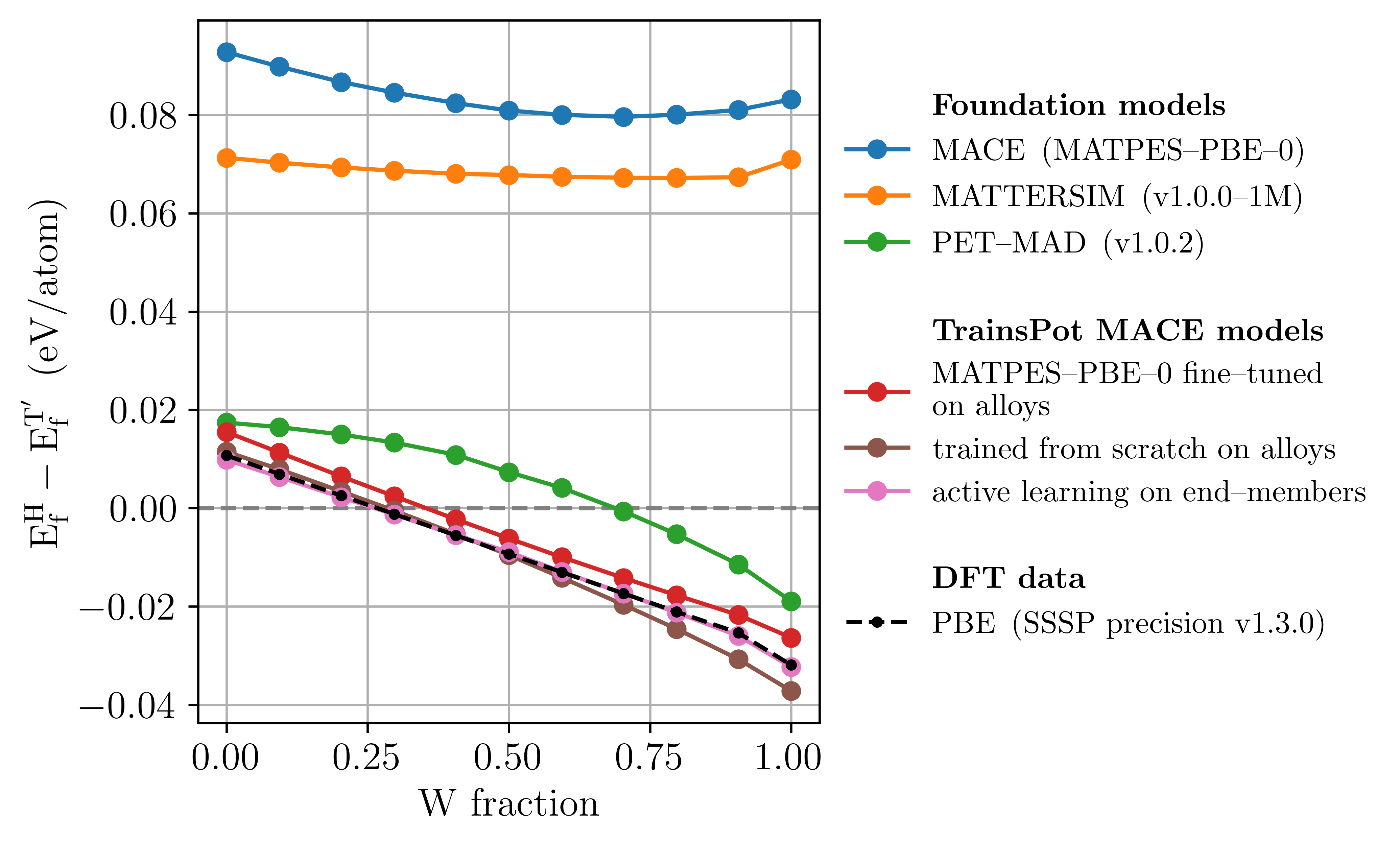}
  \caption{\label{Fig11}
    Comparison of the formation-energy difference $E_\mathrm{f}^\mathrm{H}(x) - E_\mathrm{f}^{\mathrm{T'}}(x)$ as a function of W concentration $x$ for several NNIPs, comparing foundation models with models trained or fine-tuned through AiiDA-TrainsPot.
    The MACE potentials generated with AiiDA-TrainsPot are trained on datasets produced via data augmentation starting from pure $\mathrm{MoTe_2}$ and $\mathrm{WTe_2}$ in both H and $\mathrm{T'}$ phases. The models explicitly trained or fine-tuned on alloys share the same dataset of 1687 alloy structures generated in the dataset-augmentation stage. In contrast, the active-learning runs on end-members alloys only, i.e., no clusters, surfaces, nor substitutional configurations are generated during dataset augmentation and hence are not present in the training dataset. All DFT simulations are computed with the same level of theory, except for the training sets originally employed to build the foundation models by their respective authors.
  }
\end{figure}

For each W fraction $x$, the formation energy of a given configuration is defined as:
\begin{equation}
E_{\mathrm{f}}(x)
    = E(\mathrm{W_xMo_{1-x}Te_2})
      - x\,E(\mathrm{\mathrm{WTe_2}})
      - (1-x)\,E(\mathrm{\mathrm{MoTe_2}}),
\end{equation}
where all reference energies are computed in the most stable polymorph at zero temperature (H for $\mathrm{MoTe_2}$ and $\mathrm{T'}$ for $\mathrm{WTe_2}$).
The quantity reported in Fig.~\ref{Fig11} is the formation energy difference $E_{\mathrm{f}}^{\mathrm{H}}(x) - E_{\mathrm{f}}^{\mathrm{T'}}(x)$, which directly determines the relative phase stability.

To generate validation configurations, we employ special quasirandom structures (SQS)\cite{Zunger1990} to approximate random W/Mo distributions on the metal sublattice at a given composition $x$. For each intermediate composition shown in Fig.~\ref{Fig11}, we construct and evaluate 10 independent SQS realizations to ensure statistical robustness of the estimated formation-energy differences.

As shown in Fig.~\ref{Fig11}, existing foundation models struggle with this task. \textsc{MATTERSIM} (v1.0.0--1M) and \textsc{MACE} (MATPES--PBE--0) incorrectly predict the $T'$ phase to be the ground state for all compositions, while \textsc{PET--MAD} (v1.0.2), although capturing a phase transition and quite well the formation energy difference for end-members, places the critical concentration around $x\sim0.7$\textemdash far from the DFT reference value at about $x\sim0.3$.

Then, we employ AiiDA-TrainsPot to train three additional MACE models, all starting from the same four pristine input structures (H- and $\mathrm{T'}$-phase $\mathrm{MoTe_2}$ and $\mathrm{WTe_2}$) taken from the Materials Cloud 2D database~\cite{Mounet2020}:
\begin{enumerate}
  \item Alloy model trained from scratch.
  A dataset of 1687 alloy configurations is generated by randomly sampling the W/Mo distribution in both phases during the dataset-augmentation stage, and a committee of models is trained from scratch on this dataset.

  \item Fine-tuned alloy model.
  The same alloy dataset is used to fine-tune the MACE MATPES--PBE--0 foundation model.

  \item End-member model with active learning.
  A fully independent model was trained from scratch using only configurations of the pure end-members in both phases (no alloys, clusters, surfaces, or substitutional configurations were generated during dataset augmentation). Twelve active-learning iterations are performed, yielding a final dataset of 6965 structures.
\end{enumerate}

All three models markedly outperform the existing foundation models, in particulr accurately capturing both the composition dependence of the formation-energy difference and the H-T$'$ phase-stability crossover. 
Remarkably, the best-performing model is the one trained only on end-member configurations: despite never having seen alloy structures, it correctly reproduces the formation-energy trend and predicts the phase-stability crossover close to the DFT reference value. 
This result highlights the effectiveness of active learning with calibrated committee disagreement\textemdash and, more generally, of the AiiDA-TrainsPot protocol\textemdash in sampling the relevant regions of configuration space and constructing a compact dataset that yields a highly transferable NNIP.

\section{Conclusions}
We have demonstrated that a fully-automated strategy based on data augmentation and active learning\textemdash steered by a calibrated committee-disagreement for energy, forces and stress tensor components\textemdash provides an effective way to explore the PES and to train accurate NNIPs in a data-efficient manner with minimal human intervention. For instance, a potential describing carbon allotropes has been obtained with as few as 48 initial input structures that were pulled from publicly available crystal structure databases; these prototypes have been transformed into thousands of uncorrelated, diverse and relevant configurations\textemdash all generated and calculated with no human supervision. The data augmentation is performed right at the beginning of the process to obtain about 1,000 diverse and uncorrelated training configurations: this is enough to produce sufficiently accurate and stable NNIPs that can be used for classical MD simulations, which are computationally much cheaper than AIMD and can produce additional uncorrelated configurations for further refinement of the NNIP at low cost. 

The SOREP-based dimensionality reduction (tSNE) has shown that subsequent MD-based active learning steps explore the PES through a dual mechanism: dense sampling of already-known regions and, simultaneously, exploration of entirely new basins. A compelling example is the spontaneous formation of carbon nanotubes during the MD simulations in the active learning cycle: carbon nanotubes were absent from the original dataset but become then automatically incorporated by AiiDA-TrainsPot into subsequent training iterations, suggesting the ability of the automated workflow  to find novel and quite different metastable or stable structures without prior knowledge. Indeed, the automated training strategy delivered potentials that were also able to reproduce the correct RDF of amorphous carbon.

A key aspect is the use of calibrated committee disagreement to guide the selection of new training structures. This strategy improves the efficiency and reliability of active learning, while ensuring that the model is exposed just to configurations that enhance its predictive power. Notably, the relationship between committee disagreement and actual error appears to be linear over all active learning iterations and across different properties (e.g., energies, forces, stress tensors): that support the reliable use of calibrated committee disagreement also in production simulations, i.e., when reference {\it ab initio} simulations typically cannot be performed.

Furthermore, as demonstrated for both the carbon allotropes and the $\mathrm{W_xMo_{1-x}Te_2}$ alloy benchmarks, AiiDA-TrainsPot can be effectively employed to fine-tune existing foundation models to specific applications. Fine-tuning can be performed either in a \emph{single shot} on a fixed dataset or embedded into subsequent active-learning iterations, further improving model accuracy and robustness.

It would be interesting to investigate other data-augmentation approaches\textemdash such as random structure search~\cite{AIRSS}, non-diagonal supercells~\cite{williams_prb_2015}, or generative models~\cite{luo_npj_2024}\textemdash as well as advanced exploration strategies beyond NPT and NVT MD\textemdash such introducing metadynamics~\cite{laio_pnas_2002} by interfacing AiiDA-TrainsPot to PLUMED~\cite{Bonomi2009,Tribello2014,Plumed2019}. More in general, we hope that the AiiDA's plugin system and the modular structure of AiiDA-TrainsPot will encourage and facilitate future upgrades, as well as the integration of new tools. An example would be supporting multiple NNIP backends (beyond MACE and Metatrain) and electronic structure codes (beyond \textsc{Quantum ESPRESSO}), in the spirit of previous efforts on code-agnostic common workflows for EOS and dissociation curves~\cite{Huber_npjcm_2021}. Powerful upgrades would be enabled by interfacing AiiDA-TrainsPot with existing specialized AiiDA workflows, for instance using DFT+U calculations where the Hubbard U can be automatically calculated for each configuration either with DFPT~\cite{Timrov_PRB_2018, Timrov_PRB_2021, Timrov_CPC_2022} through the AiiDA-Hubbard workflow~\cite{Bastonero_npjcm_2025} or even more efficiently through ML methods~\cite{Uhrin_npjcm_2025}.

As a side note, AiiDA-TrainsPot inherit from the AiiDA infrastructure the tracking of data-provenance graphs, enabling external validation and assessment of the published training data. This capability is crucial for public foundational models and, more broadly, for the reuse of training datasets and their corresponding NNIPs by the community.

While AiiDA-TrainsPot can operate autonomously for general-purpose NNIP development, domain experts retain full flexibility to incorporate their physical and chemical intuition in the automation strategy. The modular architecture is designed to enable full customization of all key components: initial structure selection, dataset augmentation parameters, MD simulation conditions, and computational settings for integrated codes—\textsc{Quantum ESPRESSO}, MACE, Metatrain and LAMMPS. Such level of customization allows users to balance the power of automation with the flexibility that is needed to support a wide range of applications, which require tailoring the active learning process to specific research objectives. Indeed, while AiiDA-TrainsPot  automates the entire process\textemdash traditionally long, tedious and prone to human errors\textemdash of developing NNIPs, optimal results still benefit from careful consideration of the system of interest. In other words, the selection of initial structures, augmentation strategies, and MD conditions, can\textemdash and often should\textemdash be tailored to reflect the target application and desired properties: AiiDA-TrainsPot makes that effort straightforward. On top of that, the enforcement of standardized protocols (either already established, e.g., SSSP pseudopotentials~\cite{Prandini2018} or ``fast''/``stringent'' QE protocols~\cite{Nascimento2025}, or introduced in this work) contribute to precision, reproducibility and seamless integration with future efforts in training larger models. 

AiiDA-TrainsPot democratizes the access to high-quality NNIPs tailored to the application of interest, hopefully encouraging computational scientists with limited expertise in electronic structure and ML to tackle challenging phenomena and materials, pushing the frontier of what can be simulated, understood and designed with \textit{ab initio} accuracy.

\section{Methods\label{sec:Methods}}
All calculations were performed using AiiDA-TrainsPot, \textsc{Quantum ESPRESSO}~\cite{Giannozzi2009,Giannozzi2017,Giannozzi2020} v7.3.1, MACE~\cite{MACE} v0.3.12, metatrain~\cite{metatrain} v2025.10, and LAMMPS~\cite{LAMMPS} v8Feb2023.

\subsubsection*{Carbon allotropes}
The initial dataset for carbon allotropes included 48 structures, primarily sourced from the Materials Cloud 2D and 3D databases \cite{Mounet2020,Huber2022}, with additional nanoribbons and fullerenes from the Drautz dataset~\cite{Drautz}.

For both runs, structures were replicated up to a maximum of 600 atoms and a minimum cell length of 24~\AA. A total of 80 cluster structures (up to 30 atoms each with minimum interatomic distance 1.5~\AA) were generated. Slab configurations were created with a minimum thickness of 10~\AA\ and a maximum of 600 atoms, along the (100), (110), (111), (001), (011), (010), and (101) directions. Non-periodic directions were padded with 15~\AA\ of vacuum. Vacancies (2 per structure) were created in 30\% of the structures.

For \textit{run A}, random distortions and strains were introduced with a \textit{rattle\_fraction} of 0.4, a \textit{max\_compressive\_strain} of 0.2 and a \textit{max\_tensile\_strain} of 0.2. DFT calculations used the \textit{fast} protocol \cite{Nascimento2025} for k-point grid ($\lambda = 0.30~$\AA$^{-1}$) and smearing ($\sigma_{cold}=0.0275~Ry$). MD simulations explored temperatures ranging from 0 to 5000~K and pressures from $-5$ to $5$~kbar.

For \textit{run B}, dataset augmentation parameters were optimized for near-equilibrium conditions: \textit{rattle\_fraction} was reduced to 0.3, while strain ranges were increased to \textit{max\_compressive\_strain} of 0.3 and \textit{max\_tensile\_strain} of 0.6 to better sample elastic deformations. DFT calculations employed the \textit{stringent} protocol for enhanced accuracy ($\lambda=0.1~$\AA$^{-1}$, $\sigma_{cold}=0.0125~Ry$). MD exploration was constrained to temperatures from 0 to 1000~K and pressures from $-20$ to $20$~kbar.

Both runs utilized the SSSP PBE precision library v1.3 pseudopotentials~\cite{Prandini2018,paw,pslib1}, total energy convergence threshold of $10^{-8}$~Ry, MACE training with radial cutoff of 4.5~\AA, two message-passing layers, batch size of 1, and up to 500 epochs. MD simulations were performed in NPT (for fully or partially periodic systems) or NVT (for non-periodic systems) ensembles using a 1~fs timestep and extracting trajectory frames every 1~ns. Since van der Waals interactions were not included at the DFT level, Grimme's D2 dispersion correction~\cite{Grimme2006} was enabled in MD simulations via the \textit{momb} pair style~\cite{Zhou2014} in LAMMPS. Active learning thresholds on energy, forces, and stress tensor were set to 2 meV, 50 meV/\AA, 10 meV/\AA$^3$, respectively, with a maximum of 1000 structures selected per iteration.

\subsubsection*{SACADA dataset}
DFT calculations employed both the SSSP PBE and PBEsol precision pseudopotential libraries (v1.3)~\cite{Prandini2018,paw,pslib1}, using the same computational parameters as those adopted in the active learning. For fine-tuning, the PET-MAD potential (v1.0.2)~\cite{petmad} was trained using the default hyperparameters, with a batch size of 3 and 100 training epochs.

\subsubsection*{$\mathrm{W_xMo_{1-x}Te_2}$ alloys}
The initial dataset for $\mathrm{W_xMo_{1-x}Te_2}$ alloys included 4 structures comprising the H and $\mathrm{T'}$ polymorphs of pristine $\mathrm{MoTe_2}$ and $\mathrm{WTe_2}$ monolayers, sourced from the Materials Cloud 2D database~\cite{Mounet2020}. 

During dataset augmentation, structures were replicated up to a minimum cell length of 18~\AA. The non-periodic direction was padded with 15~\AA\ of vacuum. Vacancies (2 per structure) were created in 30\% of the structures. No substitutional configurations, clusters, or surfaces were generated during dataset augmentation.
Only for the alloys dataset, random W/Mo distributions on the metal sublattice were sampled by setting in dataset augmentation stage Te as \textit{fixed\_species} and Mo and W as \textit{alloy\_species}.

The same DFT computational settings as in \textit{run B} for carbon allotropes were used also in this case, while MD simulations performed in NPT ensemble (with barostat applied only to in-plane directions) explored temperatures ranging from 0 to 1800~K and pressures from $-10$ to $10$~kbar, without additional VdW corrections.
MACE trainings were performed with radial cutoff of 6.0~\AA, two message-passing layers, batch size of 5, and up to 500 epochs, except for fine-tuning run for which same parameter of the original MACE MATPES--PBE--0 model were used.
\\

All the calculations in this work were performed on the CINECA Leonardo supercomputer, using nodes equipped with 4 NVIDIA A100 GPUs.  
Taking as reference the active-learning run used to train the MACE potential for MoTe$_2$ and WTe$_2$ (alloy benchmark in Fig.~\ref{Fig11}), where each structure contains about 108 atoms, we estimate an average computational cost of $\sim 1.2$ GPU hours per structure for DFT calculations, $\sim 25$ GPU hours for each MACE training run, and $\sim 0.1$ GPU hours for each LAMMPS MD simulation.  
After 12 active-learning iterations, resulting in a final dataset of 6965 structures, the total computational cost for training this committee of MACE potentials amounts to about 11\,000 GPU hours, 84\% of which was spent on DFT labelling.

\section*{Author Contributions}
D.B. and N.M. contributed equally. A.M. conceived the project; D.B. and N.M. developed the software and performed the simulations under the supervision of M.P. and A.M.; all authors analyzed the results and contributed to writing the manuscript.
\section*{Conflicts of interests}
There are no conflicts to declare.
\section*{Data availability}
The training datasets and  trained  models are available on the Materials Cloud Archive~\cite{Data_available}.
AiiDA-TrainsPot is available on GitHub at \href{https://github.com/aiida-trieste-developers/aiida-trains-pot}{https://github.com/aiida-trieste-developers/aiida-trains-pot}.
\section*{Acknowledgements}
All authors acknowledge support from the ICSC -- Centro Nazionale di Ricerca in High Performance Computing, Big Data and Quantum Computing, funded by European Union - NextGenerationEU (CUP Grant. No. J93C22000540006, PNRR Investimento M4.C2.1.4), and ENI through the Innovation Grant APECAR. The authors acknowledge CINECA, under CINECA-SISSA agreements, for the availability of high-performance computing resources and support. A.M. and N.M. acknowledge partial support from the European Commission through the Centre of Excellence ``MaX - Materials Design at the Exascale'' (HORIZON-EUROHPC, Grant No. 101093374).  The views and opinions expressed are solely those of the authors and do not necessarily reflect those of the European Union, nor can the European Union be held responsible for them.

\bibliography{biblio}
\bibliographystyle{rsc} 
\end{document}